\documentclass{WileyMSP-template}

\usepackage{siunitx}%
\usepackage{hyperref}

\begin{document}

\pagestyle{fancy}

\title{Spinning Living Crystals of Run-and-Tumble Particles with Environmental Feedback}

\maketitle

\author{Maks Pe\v cnik Bambi\v c,$^1$}
\author{Nuno A. M. Ara\'ujo,$^{2,3}$}
\author{and Giorgio Volpe$^{1,}$*}

\begin{affiliations}
\RaggedRight
$^1$Department of Chemistry, University College London, 20 Gordon Street, London WC1H 0AJ, United Kingdom\\

$^2$Departamento de F\'isica, Faculdade de Ci\^encias, Universidade de Lisboa, 1749-016 Lisboa, Portugal \\ 
$^3$Centro de F\'isica Te\'orica e Computacional, Faculdade de Ci\^encias, Universidade de Lisboa, 1749-016 Lisboa, Portugal \\

*Email Address: g.volpe@ucl.ac.uk\\

\end{affiliations}


\justifying
\begin{abstract}

Collective rotations are common in active matter, enhancing cohesion, transport, and mixing. They are typically attributed to chiral non-reciprocal dynamics due to intrinsic particle chirality, torque-generating interactions among units, or geometric confinement. Here, we uncover a different mechanism for rotational order in active matter where a dynamic environment coordinates the self-organization of non-chiral active particles into living crystals exhibiting sustained collective solid-like rotations. At intermediate densities, feedback from a fluctuating landscape of passive Brownian particles stabilizes large living crystals of obstacle-avoiding run-and-tumble agents. Strikingly, this environmental feedback also produces living crystals with qualitatively distinct dynamics: collective solid-like spinning emerges for particles with long persistence times approaching ballistic motion, rather than for particles moving by conventional enhanced diffusion. Beyond revealing a new route to collective rotational order in active matter, these findings  highlight the integral role of a dynamic environment in self-organization and suggest environment-mediated design principles for active materials with unconventional dynamical responses.

\end{abstract}

\section*{Introduction}

Living crystals are highly ordered, far-from-equilibrium assemblies that emerge from the collective self-organization of motile units \cite{palacci_living_2013,bechinger_active_2016}. Examples appear both in biological systems, such as bacterial colonies \cite{PhysRevLett.114.158102} and groups of animals \cite{tan_odd_2022}, and in artificial active matter, such as active colloids \cite{palacci_living_2013,ginot_aggregation-fragmentation_2018,dias_environmental_2023} and macroscopic robots \cite{rubenstein2014programmable,wang_robo-matter_2024}. By merging crystalline order with active dynamics, living crystals balance cohesion with adaptive transport of units in both biological collectives and synthetic active materials \cite{tan_odd_2022,ginot_aggregation-fragmentation_2018,karani_tuning_2019}, thus improving group performance, fitness and adaptiveness compared to their equilibrium counterparts \cite{ramaswamy_mechanics_2010,needleman_active_2017}.  Among the diverse emergent behaviors of living crystals, the collective rotations exhibited, e.g., by some spinning groups \cite{tan_odd_2022,wang_robo-matter_2024,karani_tuning_2019}, vortices \cite{caprini_self-reverting_2024,canavello_polar_2024} and swirls \cite{aubret2018targeted} can enhance cohesion \cite{Lavergne2019}, mixing \cite{ai_spontaneous_2023,guo_chirality-induced_2025}, and even yield unconventional mechanical properties, such as odd viscosity and elasticity \cite{tan_odd_2022,kole_layered_2021,soni_odd_2019,bililign_motile_2022}.

A key current challenge in active matter is to identify general minimal rules by which activity and interactions of simple units can be harnessed to engineer desired group structures and functions in active materials, particularly at scales that exceed those of the individual units' dynamics \cite{araujo2023steering}. For example, diverse collective behaviors, including swarming and flocking, have been realized in active colloids by engineering attractive, repulsive or aligning interactions through particle design \cite{Bricard2013,Granick2016}, external modulation of their propulsion mechanism \cite{khadka2018, Lavergne2019,karani_tuning_2019}, and, recently, dynamic environments \cite{dias_environmental_2023}. To date, sustained rotational order in active matter with fluid-like \cite{filella2018hydrodynamic,liebchen_collective_2017,liu_viscoelastic_2021,bricard2015emergent} or solid-like \cite{wang_robo-matter_2024,caprini_self-reverting_2024,huang2025anomalous,canavello_polar_2024} properties has instead been predominantly attributed to explicit sources of chirality and non-reciprocity at the particle level (e.g., asymmetric body shapes, chiral propulsion mechanisms) \cite{wang_robo-matter_2024,caprini_self-reverting_2024,huang2025anomalous,bililign_motile_2022}, due to torque-generating interactions among units (e.g, hydrodynamic coupling) \cite{liebchen_collective_2017,filella2018hydrodynamic}, or because of static geometric confinement (e.g., in circular wells) \cite{canavello_polar_2024,bricard2015emergent,liu_viscoelastic_2021}. In these scenarios, chirality is either hard-wired into the particles themselves or imposed by the environment in a static and externally prescribed manner, so that collective rotations are typically understood as a direct reflection of microscopic particles' asymmetry or boundary-induced constraints. 

Here, we demonstrate that a dynamic environment can induce sustained collective solid-like rotations of living crystals of non-chiral active particles, thus revealing an alternative, environment-driven route to rotational order in active matter. We consider non-chiral, obstacle-avoiding microscopic run-and-tumble agents moving within a fluctuating landscape of passive Brownian particles that act as mobile obstacles. At intermediate obstacle densities, feedback from the environment catalyzes the formation of stable living crystals with qualitatively different rotational dynamics, in which collective solid-like spinning emerges  for particles with long persistence times approaching ballistic motion but not for particles moving by conventional enhanced diffusion. These rotating crystal-like active structures are robust and persist over long times without any intrinsic particle chirality, direct torque-inducing interactions among units, or explicit static confinement.

\begin{figure}[h]
\includegraphics[width=0.6\linewidth]{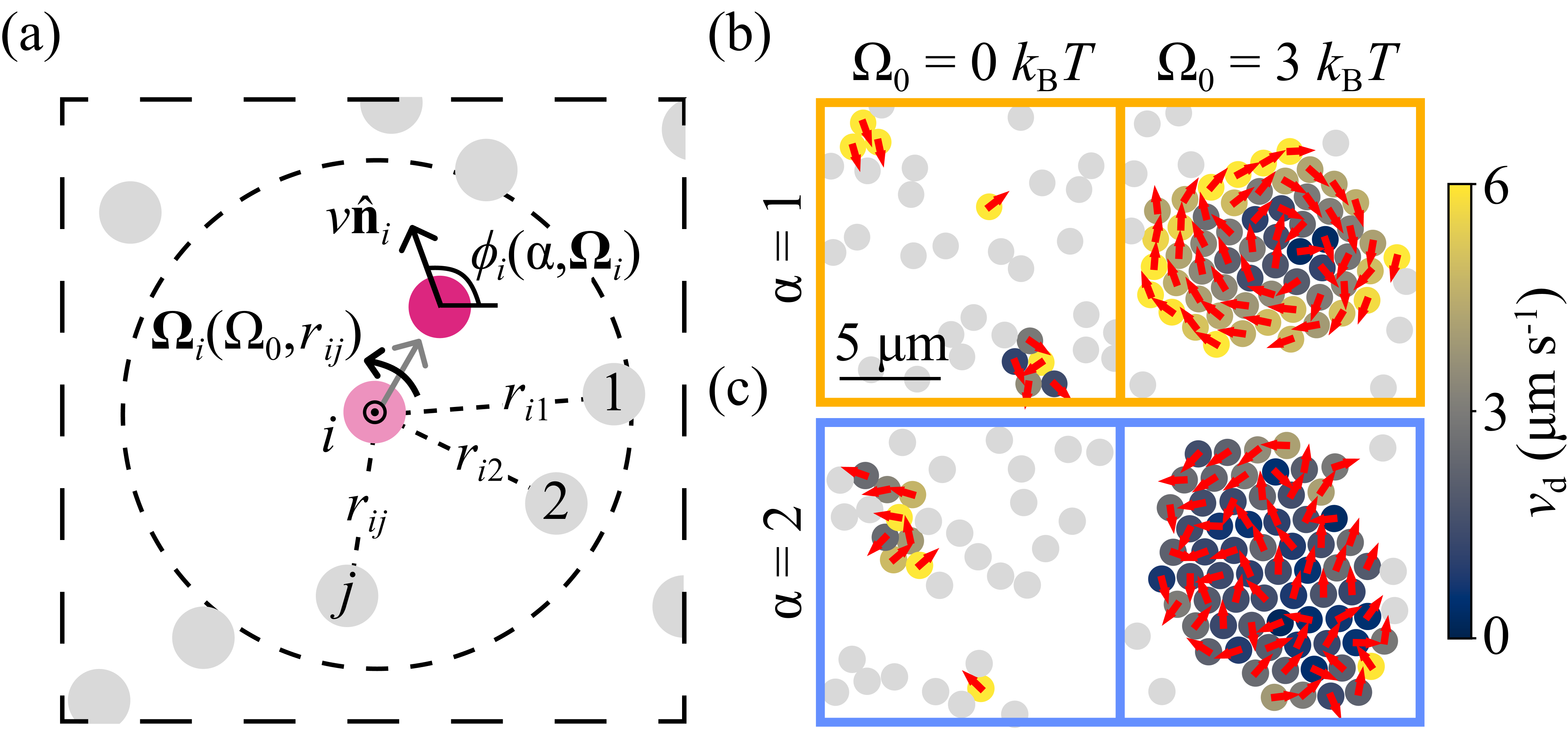}
  \caption{\textbf{Formation of spinning living crystals in crowded environments.} (a) Schematic of a microscopic run-and-tumble agent $i$ (pink) self-propelling with constant speed $v$ in a crowded environment of passive Brownian particles (gray) of same radius $R$. The angle $\phi_i$ defines the agent's self-propulsion direction ($\mathbf{v}_i = v \hat{\mathbf{n}}_i = v [\cos \phi_i, \sin \phi_i]$) as a function of the L\'evy exponent $\alpha$ and the environmental torque $\mathbf{\Omega}_i(\Omega_0,r_{ij})$ with strength $\Omega_0$ that it experiences due to the presence of the $j$-th passive particle at distance $r_{ij}$ (within a cut-off of $8R$, dashed circle). The effect of the torque (curved black arrow) is to steer the active particle away from areas of high passive density \cite{dias_environmental_2023}. (b-c) Simulation snapshots (when average crystal size stops growing at $t=750 \textrm{ s}$ from the start of the simulation, Figs. \ref{fig:clust_snaps_small} and \ref{fig:clust_snaps_big}) of (b) superdiffusive ($\alpha = 1$) and (c) normal diffusive ($\alpha = 2$) run-and-tumble agents (colored-coded for their displacement speed $v_{\rm d}=||\dot{\mathbf{x}}_i||$ with $\mathbf{x}_i$ the $i$-th particle's position) without ($\Omega_0= 0 \ k_\textrm{B}T$, left column) and with environmental torque ($\Omega_0= 3 \ k_\textrm{B}T$, right column) at a density $\rho_\textrm{p}=15\%$ of passive Brownian particles.  The environmental torque facilitates the formation of larger living crystals, whose rotational dynamics depend on $\alpha$ and, hence, on the particles' average tumbling time. The red arrows representing the unit vector  $\dot{\mathbf{x}}_i / v_{\rm d}$ in the direction of each agent's displacement show that only the superdiffusive particles form living crystals that support sustained collective solid-like spinning.}
  \label{fig:model_scheme}
\end{figure}

\section*{Results}

\subsection*{Spinning living crystals in crowded environments}

We model our active agents as run-and-tumble particles of radius $R = 0.7 \, \si{\micro\meter}$ (a size comparable to bacterial cells, such as \textit{E. coli} \cite{berg2004coli}) moving at constant speed $v$ in crowded aqueous environments of same-size passive Brownian particles at temperature $T = 298 \, \si{K}$ (Figs. \ref{fig:model_scheme},  \ref{fig:clust_snaps_small} and \ref{fig:clust_snaps_big}, Methods). The passive particles act as dynamic obstacles for the active ones. We consider a generalized description of run-and-tumble particles based on L\'evy walks, a popular stochastic model describing anomalous dynamics that deviate from Brownian motion \cite{zaburdaev_levy_2015}, as observed for many biological  \cite{berg2004coli,klages2025modelling,ariel_swarming_2015,huo_swimming_2021,waigh_heterogeneous_2023} and synthetic run-and-tumble particles \cite{karani_tuning_2019,gentili2025anomalous}. L\'evy walks feature heavy-tailed run-length distributions, where an agent moves in a straight run for a time $\tau_k$ drawn from an $\alpha$-stable L\'evy distribution $P_\alpha(\tau)$, before selecting a new random orientation $\phi$ (tumble) for the next run (Fig. \ref{fig:model_scheme}a, Methods) \cite{zaburdaev2016superdiffusive}. The average tumbling time increases by decreasing the L\'evy exponent $\alpha$. Two values define limiting dynamics in the range where the distributions $P_\alpha(\tau)$ have a finite average, $\alpha \in [2,1)$ \cite{viswanathan2011physics}: for $\alpha = 2$, $P_\alpha(\tau)$ scales exponentially, yielding normal diffusion with a linear asymptotic mean squared displacement in homogeneous environments (${\rm MSD(\Delta t)} \propto \Delta t$, Fig. \ref{fig:homo_vels}a); for $\alpha = 1$, $P_\alpha(\tau) \propto \tau^{-2}$ shows power-law scaling instead, yielding superdiffusion with a ballistic asymptotic MSD in homogeneous environments (${\rm MSD(\Delta t)} \propto \Delta t^2$,  Fig. \ref{fig:homo_vels}a). In our simulations (capped after $1000 \textrm{ s}$), the average persistence time $\langle \tau_k \rangle$ increases from $0.29 \textrm{ s}$ for normal diffusive particles ($\alpha = 2$) to $1.33 \textrm{ s}$ for superdiffusive particles ($\alpha = 1$). For comparison with known dynamics, the case $\alpha = 2$ reproduces the reference dynamics of a standard active Brownian particle of same size and speed, moving by enhanced diffusion set by its rotational Brownian motion (Fig. \ref{fig:homo_vels}a, Methods) \cite{bechinger_active_2016,volpe_simulation_2013}. 

\begin{figure}[h!]
  \includegraphics[width=0.6\linewidth]{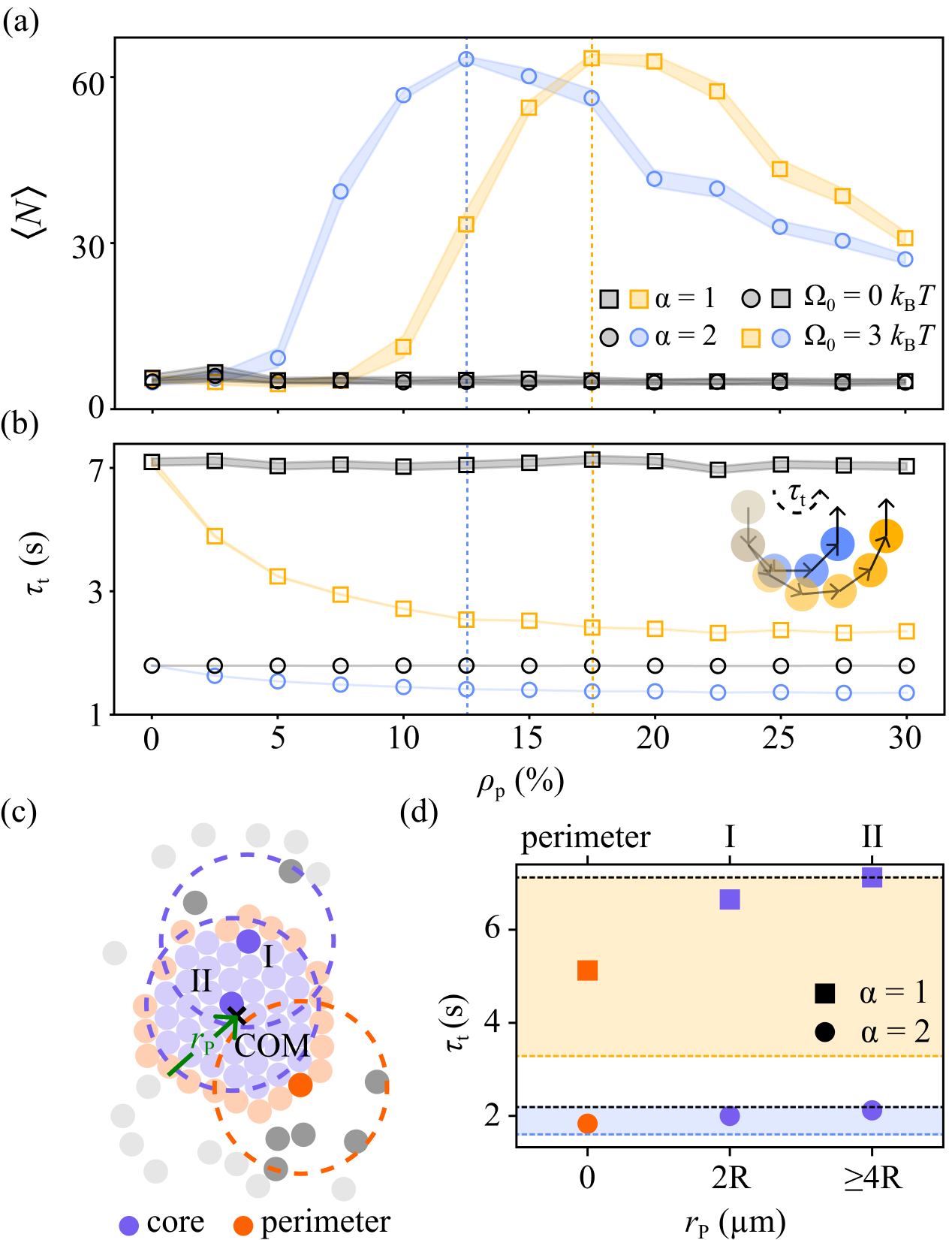}
\caption{\textbf{Formation and stabilization of large living crystals by environmental feedback.} (a) Average crystal size $\langle N \rangle$ after $1000 \textrm{ s}$ and (b) re-orientation time $\tau_\textrm{t}$ of individual active particles (before joining a crystal), defined as the average time it takes a particle to rotate by $\pi \textrm{ rad}$ (shown schematically), as a function of passive density $\rho_\textrm{p}$ for different values of the L\'evy exponent $\alpha$ and strength $\Omega_0$ of the environmental torque: $\alpha = 1$ (squares) and $\alpha = 2$ (circles) at $\Omega_0 = 0 \ k_\textrm{B}T$ (gray symbols) and $\Omega_0 = 3 \ k_\textrm{B}T$ (colored symbols). 
For $\Omega_0 \ne 0 \ k_\textrm{B}T$, (a) $\langle N \rangle$ depends non-monotonically on $\rho_\textrm{p}$, while (b) $\tau_\textrm{t}$ decreases before plateauing at $\rho_\textrm{p}\approx 17.5\%$ for $\alpha = 1$ and at $\rho_\textrm{p}\approx12.5\%$ for $\alpha = 2$, i.e., around the respective peak positions in $\langle N\rangle$ (dashed vertical lines). Data in (a-b) are averages of 60 1000-s long and 20 200-s long independent simulations, respectively. Shaded regions represent standard error. In (b), active particles only interact sterically and attractive interactions are switched off to prevent bias from crystal formation dynamics.
(c) Schematic of a living crystal in a crowded environment of passive particles (gray) highlighting perimeter (orange) and core (purple) particles. Core particles are defined as all non-perimeter particles, i.e. those with more than 4 neighbors within a $2.2R$ center-to-center distance. 
The environmental torque (Eq. \ref{eq:torque}) is strongest for perimeter particles and weakens with distance $r_{\rm P}$ (green arrow) toward the center of mass (COM, black cross) for core particles (I and II) due to less passive particles (dark gray) within their interaction range (dashed circles), (d) thus increasing $\tau_\textrm{t}$ (here at $\rho_\textrm{p} = 15\%$; other $\rho_\textrm{p}$ values in Fig. \ref{fig:tau_t_clust})  from values characteristic of individual particles with torque (colored dashed lines for $\alpha = 1$ and $\alpha = 2$) at the perimeter to those characteristic of individual particles without torque (black dashed lines) near the COM. Reference dashed lines from values in b at $\rho_\textrm{p} = 15\%$. Data from crystals with at least 20 particles. Error bars representing standard errors are smaller than the symbol size. 
}
  \label{fig:clust_size}
\end{figure}

As observed in typical experiments with active colloids \cite{palacci_living_2013,buttinoni2013dynamical,mognetti2013living,dias_environmental_2023}, our active agents also interact with each other through steric and short-range attractive interactions (Methods, Fig. \ref{fig:lj_fit}); encounters can then lead to the formation of metastable clusters known as living crystals (here defined as groups of at least two particles separated by a center-to-center distance of at most $2.2 R$ from their nearest neighbor) \cite{palacci_living_2013,ginot_aggregation-fragmentation_2018}, whose solid-like translational and rotational dynamics are consistent with previous experimental observations on standard active colloids (Fig. \ref{fig:homo_vels}b-c) \cite{ginot_aggregation-fragmentation_2018}. We consider active particles at low density ($\rho_{\rm a} = 1.5\%$, defined as fractional surface coverage) so that, in the absence of obstacles ($\rho_{\rm p} = 0\%$), encounters are sparse and living crystal formation is limited to highly dynamic small colloidal molecules (average number of particles $\langle N \rangle  \approx 5$, Figs. \ref{fig:clust_size}a and \ref{fig:clust_snaps_homogen})  \cite{schmidt2019light}. In the presence of obstacles ($\rho_{\rm p} > 0\%$), each active particle $i$, experiences both a reduced motility due to crowding (MSDs in Fig. \ref{fig:heter_vels}a) and an effective torque $\mathbf{\Omega}_i$ of strength $\Omega_0$ that steers it away from the obstacles (Fig. \ref{fig:model_scheme}a, Methods), as also observed in our previous experiments with Janus particles \cite{dias_environmental_2023}. Obstacle avoidance is indeed a common occurrence in microscopic systems of active colloids \cite{dias_environmental_2023} and bacterial cells \cite{makarchuk_enhanced_2019}, due to aligning interactions with boundaries and/or hydrodynamics \cite{das2015boundaries,simmchen2016topographical}. As seen in Fig. \ref{fig:model_scheme}b-c (and relative time sequences in Figs. \ref{fig:clust_snaps_small} and  \ref{fig:clust_snaps_big}), for $\rho_{\rm p} \ne 0\%$, living crystals can grow larger in size compared to a homogeneous environment (Figs. \ref{fig:clust_size}a and \ref{fig:clust_snaps_homogen}). For both $\alpha$ values, when $\Omega_0 \ne 0 \, k_{\rm B}T$, the largest crystals (up to a maximum of $\langle N \rangle  \approx 65$) form for intermediate values of $\rho_{\rm p}$ (Fig. \ref{fig:clust_size}a), reaching peak size at different $\rho_{\rm p}$ values for the superdiffusive and normal diffusive particles (from $\rho_{\rm p} \approx 17.5\%$ for $\alpha = 1$  to a lower density of $\rho_{\rm p} \approx 12.5\%$ for $\alpha = 2$, Fig. \ref{fig:clust_size}a). If $\Omega_0 = 0 \, k_{\rm B} T$, crystals remain small instead ($\langle N \rangle  \le 5$, comparable to the sizes for $\rho_{\rm p} = 0\%$, Figs. \ref{fig:clust_size}a and \ref{fig:clust_snaps_homogen}), displaying qualitatively similar (yet reduced due to collisions with the obstacles) translational and rotational dynamics than in homogeneous environments (Figs. \ref{fig:clust_snaps_homogen}, \ref{fig:homo_vels}b-c and \ref{fig:heter_vels}b-c). These observations highlight the critical role of the environmental torque in stabilizing the formation of large living crystals at low active densities $\rho_{\rm a}$.

Strikingly, for $\rho_{\rm p} \ne 0\%$ and $\Omega_0 \ne 0 \, k_{\rm B}T$, crystals' rotational dynamics appear to be qualitatively very different based on the particles' average persistence time (Fig. \ref{fig:model_scheme}b-c): for  $\alpha = 1$, crystals support sustained collective spinning, as highlighted by the particles' displacement velocity $v_{\rm d}$ increasing in magnitude orthogonally to the crystal's edge when moving away from its center of mass (Fig. \ref{fig:model_scheme}b); this behavior is however absent for $\alpha = 2$ (Fig. \ref{fig:model_scheme}c). Unsurprisingly, compared to the small crystals formed in homogeneous environments ($\rho_{\rm p} = 0\%$) and in crowded environments without torque ($\rho_{\rm p} \ne 0\%$, $\Omega_0 = 0 \, k_{\rm B}T$), the larger living crystals formed in crowded environments with torque ($\rho_{\rm p} \ne 0\%$, $\Omega_0 \ne 0 \, k_{\rm B}T$)) have slower translational ($\approx 75\%$) and rotational ($\approx 70\%$) dynamics on average (Fig. \ref{fig:heter_vels}b-c), yet now crystals of superdiffusive particles rotate $\approx 60\%$ faster than their diffusive counterparts (Fig. \ref{fig:heter_vels}c, inverting the trend observed for the smaller crystals) and a peak emerges in the distribution ${\rm P}(|\omega|)$ at $|\omega| = 0.35 \, {\rm rad \, s^{-1}}$, consistent with the rotation observed in Fig. \ref{fig:model_scheme}b. Hereby, we refer to these living crystals of superdiffusive particles ($\alpha = 1$) as \emph{spinning} in contrast to the diffusive rotational dynamics of standard living crystals ($\alpha = 2$) \cite{ginot_aggregation-fragmentation_2018}.

\subsection*{Stabilization of living crystals by environmental feedback}

To interpret the qualitative differences in the formation and dynamics of living crystals of superdiffusive ($\alpha = 1$) and normal diffusive ($\alpha = 2$) particles, we need to reflect on how the environmental torque alters the dynamics of the two types with respect to a homogeneous environment ($\rho_{\rm p} = 0\%$) and a torque-free scenario in a crowded environment ($\rho_{\rm p} \ne 0\%$, $\Omega_0 = 0 \, k_{\rm B}T$). Fig. \ref{fig:clust_size}b shows the re-orientation time $\tau_\textrm{t}$ (defined as the average time it takes a particle to rotate by $\pi \textrm{ rad}$) of individual particles before joining a crystal as a function of passive density $\rho_\textrm{p}$ for different values of the L\'evy exponent $\alpha$ and strength $\Omega_0$ of the environmental torque. When $\Omega_0 = 0 \ k_\textrm{B}T$, $\tau_\textrm{t}$ is roughly independent of $\rho_{\rm p}$. Although crowding reduces the translational mobility of individual particles and their explored area (Fig. \ref{fig:heter_vels}a) leading to smaller crystals with increasing $\rho_{\rm p}$ (Fig. \ref{fig:clust_size}a), the rotational dynamics of individual active particles are unaltered from those in homogeneous environments, as, without torque, the environment does not affect the particles' rotational degrees of freedom and changes in orientation are solely due to discrete tumbling events. Therefore, $\tau_\textrm{t}$ implicitly follows the different statistics of the tumbling events, giving a higher $\tau_\textrm{t}\approx7.1 \textrm{ s}$ for superdiffusive particles than for normal diffusive particles ($\tau_\textrm{t}\approx2.2 \textrm{ s}$), consistent with their respective distributions of tumbling times (Fig. \ref{fig:homo_vels}a). In the presence of an environmental torque  ($\Omega_0 \ne 0 \ k_\textrm{B}T$), reorientation results from the combined effects of tumbling and torque-induced rotation. As $\rho_{\rm p}$ increases, particles reorient more rapidly, leading to a decrease in $\tau_\textrm{t}$ that eventually saturates from intermediate densities, coinciding with the formation of the largest living crystals at $\rho_{\rm p} = 17.5\%$ for $\alpha = 1$ and at $\rho_{\rm p} = 12.5\%$ for $\alpha = 2$ (Fig. \ref{fig:clust_size}a-b). This saturation happens because particles start being caged by frequent obstacle interactions in denser environments preventing them to experience lower $\tau_\textrm{t}$. Therefore, before the peaks in Fig. \ref{fig:clust_size}a, quicker reorientation away from the obstacles with increasing $\rho_{\rm p}$ stabilizes living crystals better by aligning the particles' self-propulsion direction toward the center of mass (COM) of the forming crystal,  where the passive particle density is lowest, thereby promoting crystal growth in time through the addition of new units (Fig. \ref{fig:clust_snaps_big}). Once oriented towards the COM, particles are  less likely to escape via tumbling or Brownian motion, due to both short-range attractive interactions (Fig. \ref{fig:lj_fit}) and the stabilizing feedback introduced by the environmental torque (Fig. \ref{fig:clust_size}b). Even when escapes occur, the environmental torque limits the ability of particles to effectively distance themselves from the crystal by swiftly reorienting them towards its interior. Superdiffusive particles exhibit larger $\tau_\textrm{t}$ than normal diffusive particles and, as a consequence, require a denser environment (corresponding to a stronger effective torque) to form and stabilize the largest crystals, leading to the shift of peak crystal formation observed in Fig. \ref{fig:clust_size}a. This behavior originates from their greater average persistence time (i.e. less frequent tumbling): superdiffusive particles deflect from obstacles with larger curvature radii ($13.1 \  \si{\micro \meter}$ for $\alpha=1$ versus $7.5 \ \si{\micro \meter}$ for $\alpha=2$ at $\rho_\textrm{p}=15\%$ and $\Omega_0 = 3 \ k_\textrm{B}T$, estimated as half the particles' average displacement during $\tau_\textrm{t}$), resulting in weaker spatial localization; equally, they also have a greater likelihood of escaping crystals when pointing away from their COM compared to normal diffusive particles for similar passive densities and, hence, torque strengths (Fig. \ref{fig:clust_size}b). Past the peaks, increasing $\rho_{\rm p}$ progressively suppresses crystal formation, as reduced particle mobility due to crowding (Fig. \ref{fig:heter_vels}a) and stronger localization effects lead to a decrease in $\langle N \rangle$ and the non-monotonic trends observed in Fig. \ref{fig:clust_size}a in the presence of the environmental torque \cite{dias_environmental_2023}.

Once in a crystal, particles experience the environmental torque differently based on their position within (Figs. \ref{fig:clust_size}c-d and \ref{fig:tau_t_clust}). Since the environmental torque decays with increasing obstacle distance (Eq. \ref{eq:torque}), particles on the perimeter experience a stronger torque as they are more likely to be surrounded by passive particles within their interaction range (Figs. \ref{fig:clust_size}c-d and \ref{fig:tau_t_clust}). Conversely,  particles in the crystal core (I in Figs. \ref{fig:clust_size}c-d and \ref{fig:tau_t_clust}) experience weaker torques the further they are from the perimeter as they interact with fewer obstacles and, if they are located very close to the COM (II in Figs. \ref{fig:clust_size}c-d and \ref{fig:tau_t_clust}), may not interact with any at all. The rotational dynamics of these core particles near the COM are therefore solely determined by their tumbling statistics as if they were in a homogeneous environment (Fig. \ref{fig:clust_size}d). The closer a particle is to the perimeter, the more their rotational dynamics instead resemble those of individual particles moving in a crowded environment in the presence of the torque (albeit at an effective lower passive density due to the radial anisotropy in exposure to the obstacles, Figs. \ref{fig:clust_size}d and \ref{fig:tau_t_clust}). This gradient in radial alignment strength towards the COM of the crystal creates a steric confinement that limits the capability of the particles within the crystal core to escape, thus further stabilizing the living crystal. 

\begin{figure}[!h]
  \includegraphics[width=0.6\linewidth]{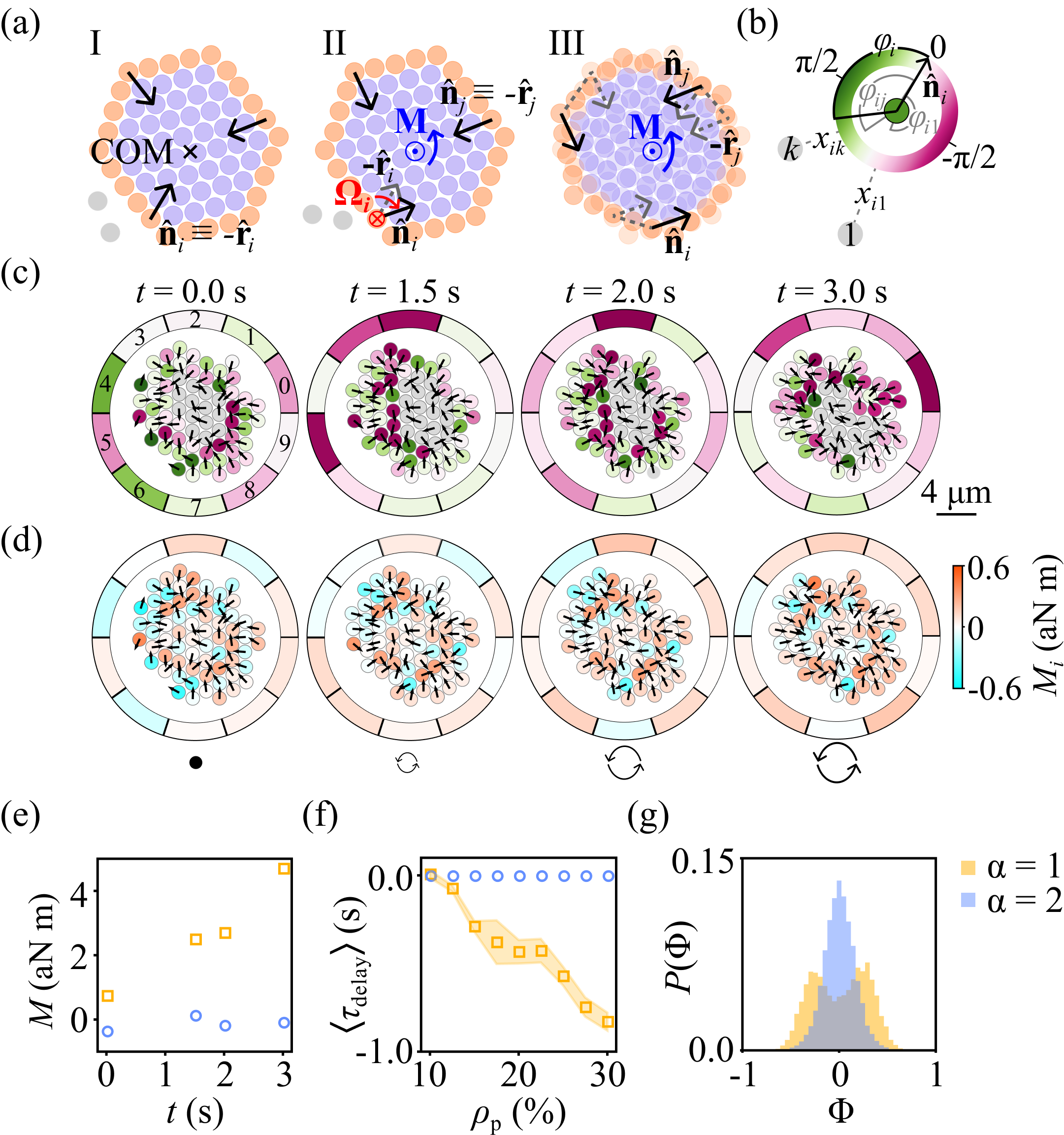}
\caption{\textbf{Living crystal spinning triggered by environmental fluctuations.} 
(a) Schematic of obstacles (gray) triggering living crystals spinning ($\alpha = 1)$: (I) as crystals form, the environmental torque anti-aligns particle $i$'s self-propulsion direction $\hat{\mathbf{n}}_i$ (black arrows) with $\hat{\mathbf{r}}_i$, the radial unit vector connecting the crystal's center of mass (COM) to the particle ($\hat{\mathbf{n}}_i \equiv -\hat{\mathbf{r}}_i$); (II) local torques $\mathbf{\Omega}_i$ due to uneven obstacle distributions reorient $\hat{\mathbf{n}}_i$ away from $-\hat{\mathbf{r}}_i$ ($\hat{\mathbf{n}}_i \not\equiv -\hat{\mathbf{r}}_i$) for a subset of particles (dashed to solid arrows), producing a torque $\mathbf{M}$ on the crystal (Eq. \ref{eq:torque}) that makes it spin; (III) individual self-propulsion directions remain fixed during rotation (dashed to solid arrows joined by dashed lines) except during tumbles, so that $\hat{\mathbf{n}}_j \not\equiv -\hat{\mathbf{r}}_j$ also for other particles, reinforcing spinning. (b) Average weighted direction of nearby obstacles (gray) for a run-and-tumble particle $i$ (center), relative to $\hat{\mathbf{n}}_i$, given by the weighted average angle $\varphi_{i}$ (black arc) of the angles $\varphi_{ik}$ between $\hat{\mathbf{n}}_i$ and the vectors connecting particle $i$ to obstacle $k$, weighted by the inverse separation distance $x^{-1}_{ik}$ (Methods). Particles are color-coded pink or green (as in this example) if obstacles lie on average to the right or left of $\hat{\mathbf{n}}_i$, inducing a counterclockwise or clockwise rotation respectively; white represents cases of no torque due to obstacles lying directly in front ($\varphi_{i}=0$) or at the back ($\varphi_{i}=\pi$) of an active particle on average. (c-d) Exemplary sequence of crystal starting to spin showing each particle $i$'s (c) $\varphi_{i}$ and (d) contribution $M_i$ to the signed scalar torque $M$ on the crystal (Methods). Black arrows represent $\hat{\mathbf{n}}_i$. Outer annular arcs (from 0 to 9) show the sector average of the respective quantity. Symbols below d represent no spinning (dot) and increasing rotation (circular arrows). Core gray particles in c do not interact with obstacles. (e) Time evolution of $M$ for the sequences in d ($\alpha = 1$, yellow) and Fig. \ref{fig:a2_snaps}b ($\alpha = 2$, blue). 
(f) Average delay $\langle \tau_\textrm{delay} \rangle$ between the time evolution of the average obstacle direction for the crystal $\varphi=\sum^{N}_i\varphi_i$ (with $N$ the number of particles in the crystal) and that of $M$ (Methods, Fig. \ref{fig:corr}). Shaded region represents standard error. (g) Probability distributions of the normalized signed scalar torque $\Phi$ (Methods) for $\alpha = 1$ and $\alpha = 2$. Data in f and g from over 60 1000-s long simulations for each $\alpha$. 
 }
  \label{fig:clust_rot}
\end{figure}

\subsection*{Role of environmental fluctuations in starting crystal spinning}

As observed in Figs. \ref{fig:model_scheme}b-c, \ref{fig:clust_snaps_big} and \ref{fig:heter_vels}b-c, living crystals formed by superdiffusive particles can exhibit sustained spinning, unlike standard diffusive ones. As spinning does not occur in the absence of obstacles (Figs. \ref{fig:clust_snaps_homogen} and \ref{fig:heter_vels}), it must originate from an interplay between the environmental torque and the dynamics of the superdiffusive particles in a crystal. To better understand the origin of these rotations, we can consider an exemplary time sequence of a spinning living crystal for $\alpha = 1$ (Fig. \ref{fig:clust_rot}). 

Initially, as crystals form, the stabilizing effect of the environmental torque (Fig. \ref{fig:clust_size}d) forces particles to point toward the region of lowest obstacle density (typically the crystal's center of mass, COM) so that the unit vector $\hat{\mathbf{n}}_i$ of each particle $i$'s self-propulsion direction tends to anti-align with the radial unit vector $\hat{\mathbf{r}}_i$ connecting the crystal's COM to the particle ($\hat{\mathbf{n}}_i \equiv -\hat{\mathbf{r}}_i$, Fig. \ref{fig:clust_rot}a I). This tendency can be visualized by estimating an average weighted direction $\varphi_i \in (-\pi, \pi]$ of nearby obstacles for each particle $i$ in the crystal relative to its $\hat{\mathbf{n}}_i$ (Fig. \ref{fig:clust_rot}b, Methods). This quantity provides an estimate of the strength and direction of the reorientation experienced by each particle based on the number and distance of surrounding obstacles: depending on whether $\varphi_i > 0$ (green in Fig. \ref{fig:clust_rot}b) or $\varphi_i < 0$ (pink in Fig. \ref{fig:clust_rot}b), each particle experiences a respectively clockwise or counterclockwise reorientation (due to the local environmental torque $\mathbf{\Omega}_i$) towards the crystal's COM (Fig. \ref{fig:clust_rot}c-d). At time $t = 0  \,{\rm s}$, $\varphi_i$ near the crystal's perimeter exhibits an almost even alternation of negative and positive values (outer annular arcs in Fig. \ref{fig:clust_rot}c). As a consequence, the net torque $\mathbf{M} = \sum_{i}{\mathbf{M}}_i$ on the crystal exerted by each particle $i$ through its contributing torque $ \mathbf{M}_i = \gamma_{\rm t} v ({\mathbf{r}}_i\times\hat{\mathbf{n}}_i)$ is almost null (almost even alternation of positive and negative $M_i$ values, outer annular arcs in Fig. \ref{fig:clust_rot}d) and the rotation of the crystal is negligible (Fig. \ref{fig:clust_rot}e). 

Due to their Brownian nature, however, the density of obstacles around the crystal fluctuates (Fig. \ref{fig:clust_rot}a II) and can cause a subset of active particles in the crystal to re-orient away from its COM and align more tangentially to the crystal's perimeter, thus creating a local pocket of alignment that can reinforce the crystal's tendency to spin in a given direction. For example, in Fig. \ref{fig:clust_rot}c-d at $t = 1.5 \, {\rm s}$, several particles within the sectors corresponding to the outer arcs 5 and 6 (and 2 and 3 to a lesser extent) start to align more perpendicular to $\hat{\mathbf{r}}_i$ in a direction compatible with a counterclockwise rotation of the whole crystal, as also highlighted by the normalized scalar torque $\Phi_i$ in Fig. \ref{fig:rot_order}a becoming more positive in those sectors (Methods). By gaining this perpendicular component in their self-propulsion relative to $\hat{\mathbf{r}}_i$ (Fig. \ref{fig:clust_rot}a II), the cumulative effect of the torques $\mathbf{M}_i$ exerted by these particles produces a net torque $\mathbf{M}$ on the crystal and sets it in a counterclockwise solid-like rotation as more outer arcs (i.e., 6) in Fig. \ref{fig:clust_rot}d starts turning red (positive $M$) compared to $t = 0 \, {\rm s}$ and any blue (negative $M$) arc left (e.g., 1, 3 and 4) becomes less intense.  Different from a true solid body's rotation, each $\hat{\mathbf{n}}_i$ remains fixed (Fig. \ref{fig:clust_rot}a III) except during tumbles, so that the spinning of the crystal causes the self-propulsion direction of more particles to align away from the COM and more tangentially to the crystal edge, reinforcing the rotation of the crystal (more outer arcs turn to positive values in Figs. \ref{fig:clust_rot}d and \ref{fig:rot_order}a from $t = 0 \, {\rm s}$ to $t = 3 \, {\rm s}$) by strengthening $\mathbf{M}$ (Fig. \ref{fig:clust_rot}e) until virtually all particles rotate cooperatively. 

At lower obstacle densities ($\rho_{\rm p} \lesssim 17.5 \%$, below the peak in Fig. \ref{fig:clust_size}a), quick rotation of the crystal can cause the particles' self-propulsion direction $\hat{\mathbf{n}}_i$ to eventually align parallel to $\hat{\mathbf{r}}_i$, thus pointing away from the crystal's COM and leading to its rotation-induced fragmentation (Fig. \ref{fig:clust_break}). When the obstacle density is large enough, however, this tendency is counterbalanced by the environmental torque from the surrounding obstacles that tends to re-align $\hat{\mathbf{n}}_i$ towards the COM and antiparallel to $\hat{\mathbf{r}}_i$ (more outer arcs turn magenta in Fig. \ref{fig:clust_rot}c from $t = 0 \, {\rm s}$ to $t = 3 \, {\rm s}$), thus maintaining most particles $\hat{\mathbf{n}}_i$ almost tangential to the crystal's edge on average and capable of sustaining its rotation over time (Fig. \ref{fig:rot_order}a). Tumbling events, being more frequent for $\alpha = 2$, can disrupt this balance for standard diffusive living crystals preventing their sustained rotation ($M \approx 0$ in Fig. \ref{fig:clust_rot}e, Figs. \ref{fig:a2_snaps} and \ref{fig:rot_order}b). In contrast,  superdiffusive particles have a longer persistence time with less frequent tumble-induced reorientation events (Fig. \ref{fig:homo_vels}a), thus enabling crystals' rotations to persist.

Beyond this exemplary sequence, the typically negative values of the delay $\tau_\textrm{delay}$ (Figs. \ref{fig:clust_rot}f and \ref{fig:corr}) extracted by correlating the time evolution of the average direction $\varphi=\sum_i\varphi_i$ of nearby obstacles and that of $M$ for a crystal (Methods) confirm and generalize these observations to other crystals, highlighting how changes in $\varphi$ (i.e., in the density of obstacles around crystals) precede crystal spinning events for $\alpha =1$, while the same is not observed for $\alpha = 2$ (Figs. \ref{fig:clust_rot}f and \ref{fig:corr}). In fact, for $\alpha = 2$, $\langle \tau_\textrm{delay} \rangle = 0$ along with a faster decay of the correlation (Fig. \ref{fig:corr}e) points to an instantaneous correlation between the two signals that quickly fades in time because of the more frequent tumbling events. As $\rho_\textrm{p}$ increases (Fig. \ref{fig:clust_rot}f), denser and more even distributions of obstacles around the crystals mean that stronger environmental torques are at play orienting each particle towards the COM quicker. Fluctuations in obstacle density therefore produce weaker and slower reorientation of the superdiffusive particles ($\alpha = 1$) away from the crystal's COM that could promote a rotation, thus increasing the negative delay $\langle\tau_\textrm{delay}\rangle$ at higher passive densities (Fig. \ref{fig:clust_rot}f).

Fig. \ref{fig:clust_rot}g shows the probability distribution of the normalized signed scalar torque $\Phi$ (Eq. \ref{eq:rot_align}) as a crystal-size independent measure of the rotational order exhibited by different crystals for $\alpha = 1$ and $\alpha = 2$ (Methods): for $\alpha = 1$, $\Phi$ follows a symmetric bimodal distribution (peaked at $\Phi=\pm0.35$), confirming that particles are most likely to be found aligned at a given angle relative to the crystal's COM and that crystals rotate in either direction with equal probabilities; conversely, $\Phi$ follows a Gaussian distribution for $\alpha = 2$, consistent with the distribution of the angular velocity $\omega$ in Fig. \ref{fig:heter_vels}c and earlier experimental studies assuming randomly oriented particles within crystals \cite{ginot_aggregation-fragmentation_2018}.

\subsection*{Synergy between perimeter and core particles in spinning living crystals}

\begin{figure}[!h]
  \includegraphics[width=0.75\linewidth]{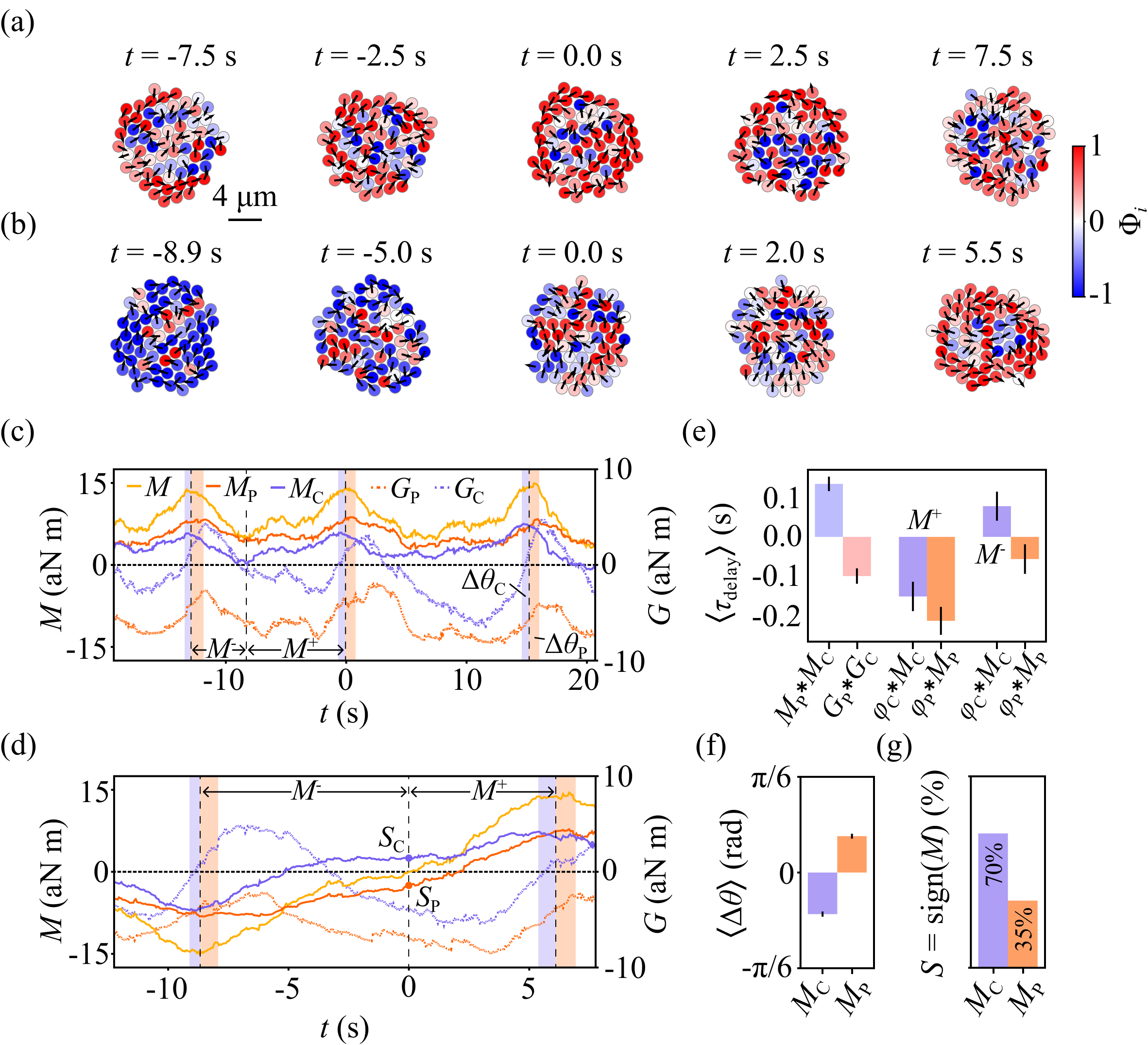}
\caption{\textbf{Rotational dynamics of spinning living crystals.} (a-b) Normalized signed scalar torque $\Phi_i$ for each particle $i$ in exemplary sequences of spinning living crystals at $\rho_{\rm p} = 15\%$ that (a) maintain spinning direction or (b) reverse it after coming to a halt at around $t = 0 \, {\rm s}$. Black arrows represent $\hat{\mathbf{n}}_i$. (c-d) Radial moments $G_{\rm P}$ and $G_{\rm C}$ of perimeter (dashed orange line) and core (dashed purple line) particles and signed scalar torque $M$ (solid yellow line) acting on the crystals in (c) a and (d) b, highlighting contributions from perimeter ($M_\textrm{P}$, solid orange line) and core ($M_\textrm{C}$, solid purple line) particles. Shaded vertical regions highlight instantaneous phase shifts $\Delta\theta_\textrm{P}$ (orange) between $M$ and $M_\textrm{P}$ and $\Delta\theta_\textrm{C}$ (purple) between $M$ and $M_\textrm{C}$ (Methods). $M^-$ and $M^+$ represent time intervals when the crystal's rotation is slowing down (maximum to minimum of $M$) or speeding up $M^+$ (minimum to maximum of $M$), obtained from the stationary points of the time derivative of $M$. In d, $S_\textrm{P}$ and $S_\textrm{C}$ are the signs of $M_\textrm{P}$ and $M_\textrm{C}$ when the crystal reverses rotation. (e) Average delays $\langle \tau_{M_\textrm{P}*M_\textrm{C}} \rangle$, $\langle \tau_{G_\textrm{P}*G_\textrm{C}} \rangle$, $\langle \tau^+_{\varphi_\textrm{C}*M_\textrm{C}} \rangle$, $\langle \tau^+_{\varphi_\textrm{P}*M_\textrm{P}} \rangle$, $\langle \tau^-_{\varphi_\textrm{C}*M_\textrm{C}} \rangle$ and $\langle \tau^-_{\varphi_\textrm{P}*M_\textrm{P}} \rangle$ obtained from the correlations $M_\textrm{P}*M_\textrm{C}$ and $G_\textrm{P}*G_\textrm{C}$ as well as $\varphi_\textrm{C}*M_\textrm{C}$ and $\varphi_\textrm{P}*M_\textrm{P}$ separately during intervals $M^+$ and $M^-$ (Fig. \ref{fig:corr}f-g). (f) Average phase shifts $\langle \Delta\theta_\textrm{C} \rangle$ and $\langle \Delta\theta_\textrm{P} \rangle$ of the instantaneous phase shifts $\Delta\theta_\textrm{C}$ and $\Delta\theta_\textrm{P}$ (highlighted in c and d, Methods). (g) Percentage of reversal events for which the signs $S_\textrm{C}$ and $S_\textrm{P}$ as in d match that of $M$ following the reversal ($S = \textrm{sign}(M)$). Data in e-g are averaged across 60 simulations at $\rho_\textrm{p}=15\%$ with vertical lines indicating standard error.}
  \label{fig:clust_rev}
\end{figure}

At intermediate passive densities ($17.5 \% \lesssim \rho_{\rm p} \lesssim 30 \%$), living crystals for $\alpha = 1$ can sustain spinning in the same direction for long stretches of time (up to $\approx 70 \textrm{ s}$, Fig. \ref{fig:time_rev}) due to the stabilizing effect of the environmental torque (Fig. 2). Crystals continue to rotate through pseudo-periodic cycles (with an average period $\tau_{\rm cyc} \approx 11.5 \, {\rm s}$, Fig. \ref{fig:time_rev}) caused by an interplay between the action of perimeter and core particles in response to the environmental torque (Fig. \ref{fig:clust_rev}).
At the end of each cycle when the crystal does not fragment (Fig. \ref{fig:clust_break}), two outcomes are possible: either it restarts spinning in the same direction as the previous cycle (Figs. \ref{fig:clust_rev}a and \ref{fig:viril_dyn}a) or in the opposite direction (Figs. \ref{fig:clust_rev}b and \ref{fig:viril_dyn}b). In Fig. \ref{fig:clust_rev}a, the normalized scalar torque $\Phi_i$ indeed maintains a positive value for most particles (indicating a counterclockwise rotation, also confirmed by the positive values of $M$ in Fig. \ref{fig:clust_rev}c), while it swaps sign through $t = 0 \, {\rm s}$ from negative (clockwise rotation) to positive (counterclockwise rotation) in Fig. \ref{fig:clust_rev}b, as also confirmed by a change in sign of $M$ in Fig. \ref{fig:clust_rev}d. 

As noted earlier, during a rotational cycle, particles' self-propulsion directions drifts outwards away from the crystal's COM against the stabilizing effect of the environmental torque that reorients them in the opposite direction (Fig. \ref{fig:clust_rot}). We can better quantify this tendency with the radial moment $G$ (Fig. \ref{fig:viril_dyn}, Methods), which measures the overall tendency of the living crystal to expansion (positive $G$) or contraction (negative $G$) depending on whether particles self-propel away or toward the COM on average. While most perimeter particles display an oscillating negative radial moment $G_{\rm P}$ for both sequences in Fig. \ref{fig:clust_rev} due to the stabilizing effect of the environmental torque (Figs. \ref{fig:clust_rev}c-d and \ref{fig:viril_dyn}), core particles behave differently since the torque is weaker closer to the COM (Figs. \ref{fig:clust_size}d and \ref{fig:tau_t_clust}). At the peak of a spinning cycle (maxima of $|M|$ in Fig. \ref{fig:clust_rev}c-d), these particles display a transition from a negative to a positive radial moment $G_{\rm C}$ in correspondence to the crystal rotation starting to slow down ($M$ decreasing in magnitude, e.g. at $t = 0 \, {\rm s}$ in Fig. \ref{fig:clust_rev}a,c and $t = -8.9 \, {\rm s}$ in Fig. \ref{fig:clust_rev}b,d). The opposite transition occurs when a new spinning cycle starts ($M$ increasing in magnitude, e.g. at $t = -7.5 \, {\rm s}$ in Fig. \ref{fig:clust_rev}a,c and $t = 0 \, {\rm s}$ in Fig. \ref{fig:clust_rev}b,d). The core particles therefore play a critical role in the crystal's rotational dynamics: for the perimeter particles to reinforce the crystal's rotation (positive $\langle \tau_{M_\textrm{P}*M_\textrm{C}} \rangle$, Fig. \ref{fig:clust_rev}e), there is a prerequisite requirement for the core particles to be favorably aligned with the new spinning direction (negative $\langle \tau_{G_\textrm{P}*G_\textrm{C}} \rangle$, Fig. \ref{fig:clust_rev}e); similarly, the rotation starts to slow down because of the spinning-induced re-alignment of the core particles away from the COM (Figs. \ref{fig:clust_rev}c-d and \ref{fig:viril_dyn}).

Initially, just before a crystal starts spinning in a given direction ($M^+$ interval, Fig. \ref{fig:clust_rev}c-d), core particles that are on average aligned towards the COM (negative $G_{\rm C}$ values, Figs. \ref{fig:clust_rev}c-d and \ref{fig:viril_dyn}) start pointing perpendicular to it, $G_{\rm C}$ reduces in magnitude towards null values (perpendicular alignment to the COM), and core particles start exerting a small torque on the crystal in response to fluctuations in the average direction $\varphi_\textrm{C}$ of their nearby obstacles (negative $\langle \tau_{\varphi_\textrm{C}*M_\textrm{C}} \rangle$, Figs. \ref{fig:clust_rev}e and \ref{fig:corr}f). Being closer to the COM, however, the torque $M_{\rm C}$ from the core is small compared to the one that the perimeter can exert ($M_{\rm P}$, Fig. \ref{fig:clust_rev}c-d). $M_{\rm C}$ alone is not significant enough to properly set the crystal in rotation until the perimeter particles follow and also align in the same direction in response to fluctuations in the average direction $\varphi_\textrm{P}$ of their nearby obstacles ($\langle \tau_{\varphi_\textrm{P}*M_\textrm{P}} \rangle < \langle \tau_{\varphi_\textrm{C}*M_\textrm{C}} \rangle$, Figs. \ref{fig:clust_rev}e and \ref{fig:corr}f), thus driving the reinforcement of the crystal's rotation (positive $\langle \tau_{M_\textrm{P}*M_\textrm{C}} \rangle$, Fig. \ref{fig:clust_rev}e). As long as $G_{\rm C} < 0$ (point at which the core particles tend to self-propel  tangentially to the crystal's edge on average), this re-alignment indeed reinforces the crystal's rotation and leads to the crystal spinning faster (higher torques in Fig. \ref{fig:clust_rev}c-d).

As soon as $G_{\rm C} > 0$ ($M^-$ interval, Fig. \ref{fig:clust_rev}c-d), core particles start aligning in the radial direction away from the COM as a consequence of the crystal's faster rotation and $M_\textrm{C}$ starts to decrease (Fig. \ref{fig:clust_rev}c-d). Differently from a fragmentation scenario (Fig. \ref{fig:clust_break}), perimeter particles instead hardly align in this direction because of the stronger stabilizing effect of the environmental torque (typical $G_{\rm P} < 0$, Figs. \ref{fig:clust_rev}c-d and \ref{fig:viril_dyn}), preventing the crystal from deforming and potentially fragmenting. The decrease in $M_\textrm{C}$ precedes that in $M_\textrm{P}$ as shown by $M_\textrm{C}$ and $M_\textrm{P}$ respectively preceding and following the total signed scalar torque $M$ (Fig. \ref{fig:clust_rev}f). Unlike perimeter particles (slightly negative $\langle \tau_{\varphi_\textrm{P}*M_\textrm{P}} \rangle$, Figs. \ref{fig:clust_rev}e and \ref{fig:corr}g), the reduction in $M_\textrm{C}$ is purely driven by the crystal rotation rather than by the environment (positive $\langle \tau_{\varphi_\textrm{C}*M_\textrm{C}} \rangle$, Figs. \ref{fig:clust_rev}e and \ref{fig:corr}g). As the crystal starts to slow down because of the decrease in $M_\textrm{C}$ and subsequent decrease in $M_\textrm{P}$, the environmental torque starts dominating the particles' dynamics, reorienting both perimeter and core particles back towards the COM ($G_{\rm C}$ and $G_{\rm P}$ turning towards more negative values, Fig. \ref{fig:clust_rev}c-d).

At the end of a cycle, a new one begins with a direction defined by the core particles. Whether a change in direction takes place or not depends on these particles reversing (Fig. \ref{fig:clust_rev}b,d) or maintaining (Fig. \ref{fig:clust_rev}a,c) the orientation of the previous cycle in response to environmental fluctuations (Fig. \ref{fig:clust_rev}e). In fact, in $70 \%$ of the spinning reversal events, at the time when the total torque $M$ on the crystal is null ($t = 0 \, {\rm s}$, Fig. \ref{fig:clust_rev}d), the sign $S_{\rm C}$ of $M_{\rm C}$ is the same as the sign of $M$ in the upcoming cycle, unlike $M_{\rm P}$ which typically has the opposite sign (Fig. \ref{fig:clust_rev}g).

\section*{Conclusion}

In summary, our simulations reveal an alternative mechanism for collective rotational order in active matter. A dynamic environment coordinates and stabilizes the emergence of living crystals of run‑and‑tumble particles that can sustain solid‑like spinning for extended stretches of time. As observed in experiments with active colloids \cite{das2015boundaries,simmchen2016topographical,dias_environmental_2023}, the physical mechanism behind the emergence of these collective dynamics is an effective torque that steers particles away from Brownian obstacles in the environment. This environment‑mediated mechanism is particularly effective for particles with long persistence times approaching ballistic motion. Motivated by increasing experimental evidence \cite{berg2004coli,klages2025modelling,ariel_swarming_2015,huo_swimming_2021,waigh_heterogeneous_2023,karani_tuning_2019,gentili2025anomalous}, our simulations adopted a generalized representation of run-and-tumble particles based on the uniform model of L\'evy walks \cite{zaburdaev_levy_2015}, but alternative statistics influencing the  average persistent time of the particles can be expected to lead to comparable results \cite{bechinger_active_2016}. A key element of the emerging rotational behavior lies in the internal structure of the spinning living crystals. The core particles set the preferred spinning direction in response to environmental fluctuations: stable rotation requires them to align favorably with the emerging collective rotation. Once the alignment of the core is established, perimeter particles can reinforce and sustain the rotation. Conversely, as spinning progresses, the induced reorientation of the core away from the center of mass gradually weakens this synergy between core and perimeter particles, ultimately slowing down the rotational motion. Thus, the interplay between both type of particles determines not only the onset but also the regulation of the collective spinning state. More broadly, our findings highlight the central role of dynamic environments in shaping self‑organization in decentralized active matter systems \cite{araujo2023steering}. Fluctuating environmental features can coordinate collective rotational motion even in the absence of intrinsic particle chirality, torque-generating interactions among units, or static geometric confinement. This suggests new design principles for active materials and artificial collectives in complex dynamic settings, where environment‑mediated feedback can be harnessed to program unconventional dynamical responses in systems of units with limited information-processing capabilities, including in the design of artificial swarm intelligence, ant-colony optimization algorithms, neuromorphic computers and crowd management control tools \cite{volpe2025roadmap}.

\section*{Methods}
\subsection*{Particle-based simulations}
Our numerical model consists of $N_\textrm{a}$ active and $N_\textrm{p}$ passive particles of radius $R = 0.7 \, \si{\micro\meter}$ moving in a square domain of side $L=120R$ with periodic boundary conditions. Every simulation begins with all particles randomly distributed without overlap within the simulation box. We assume the particles in water (viscosity $\eta = 0.001 \ \si{Pa \, s }$) at temperature $T = 298 \, \si{K}$. The density (defined as fractional surface coverage) of either species can be calculated as $\rho_\textrm{a,p}=\frac{N_\textrm{a,p}\pi R^2}{L^2}$, where ${\rm a}$ and ${\rm p}$ stand for active and passive, respectively. In our simulations, $\rho_{\rm a} = 1.5 \%$ and $\rho_{\rm p} \in [0\%, 30\%]$. The dilute regime for active particles makes encounters rare in the absence of obstacles, while low and intermediate values of passive densities (i.e., $\le 30 \%$) exclude regimes where caging events would inhibit the emergence of living crystals \cite{dias_environmental_2023}.  The passive particles are standard Brownian particles with translational and rotational diffusion coefficients given by $D_\textrm{t}=\frac{k_\textrm{B}T}{6\pi\eta R}$ and $D_\textrm{r}=\frac{k_\textrm{B}T}{8\pi\eta R^3}$, respectively ($k_\textrm{B}$ being the Boltzmann constant). Beyond being subject to translational Brownian dynamics, the active particles also perform run-and-tumble motion (according to the uniform model of L\'evy walks \cite{zaburdaev_levy_2015}) with a constant speed $v = 6 \ \si{\micro\meter \, s}^{-1}$ (a mid-range speed for several Janus particle designs \cite{bechinger_active_2016}), where the duration $\tau_k$ of each run (with $k$ an integer) is drawn from an $\alpha$-stable L\'evy distribution $P_\alpha(\tau)$ \cite{garbaczewski_computer_1995,chambers_method_1976} before a random change of the particle's orientation $\phi$ takes place and a new run starts (Figs. \ref{fig:model_scheme} and \ref{fig:homo_vels}a). We focus on the two limiting cases for $\alpha = 2$ and for $\alpha = 1$, where, asymptotically (as $\tau \rightarrow\infty$), the run durations (and, hence, the run lengths) decay exponentially when $\alpha = 2$ and follow a power law $\propto\tau^{-2}$ when $\alpha = 1$ (Figure \ref{fig:homo_vels}a). In our simulations, the distributions are scaled so that the MSD of the normal diffusive particles ($\alpha = 2$) matches that of a standard active Brownian particle of same size and self-propulsion speed (Fig. \ref{fig:homo_vels}a, inset) \cite{bechinger_active_2016}.

The trajectories of both active ($v \ne 0$) and passive ($v = 0$) particles are obtained by integrating the following overdamped equations using the Euler integration scheme with a time step of $\Delta\tau = 10 \ \si{\micro s}$ \cite{volpe_simulation_2013}:

\begin{subequations}\label{eq:motion}
    \begin{align}
        \dot{\mathbf{x}}_i &= -\frac{1}{\gamma_{\rm t}}\nabla_\mathbf{x}V_i+v\hat{\mathbf{n}}_i + \sqrt{2D_\textrm{t}}\boldsymbol{\xi}_i \ , \\
        \dot{\phi}_i &= \frac{1}{\gamma_\textrm{r}}\Omega_i+\varsigma_i \ ,
    \end{align}
\end{subequations}

\noindent
where $\mathbf{x}_i$ and $\hat{\mathbf{n}}_i=[\cos(\phi_i), \sin(\phi_i)]$ are the position and motion direction for particle $i$, $\gamma_{\rm t}=6 \pi \eta R$ is its translational friction coefficient, and $\boldsymbol{\xi}_i=[\xi_i^x, \xi_i^y]$ with $\xi_i^x$ and $\xi_i^y$ being independent white noise terms of unit variance and zero mean. The orientations $\phi_i$ of the active particles depend on two terms: the signed scalar $\Omega_i$ of an effective torque  due to the interaction of agent $i$ with the surrounding passive particles ($\gamma_\textrm{r}=8 \pi \eta R^3$ is the rotational friction coefficient) and a stochastic reorientation term $\varsigma_i = \frac{\pi}{6}\xi_i^\phi\delta(t-\tau_k)$, where $\xi_i^\phi$ is white noise of unit variance and zero mean defining the tumbling angles at times $\tau_k$. The motion of the passive particles instead is only determined by Eq. \ref{eq:motion}a (setting $v = 0$), as the rotational degree of freedom does not influence their translational dynamics in the overdamped regime. 

The interaction between particles is implemented with a Lennard-Jones potential given by 

\begin{equation}\label{eq:lj}
    V_i=\sum_j V_{ij}(r_{ij}) = \sum_j 4\epsilon_{\textrm{LJ}}\left[\left(\frac{\sigma_{\textrm{LJ}}}{r_{ij}}\right)^{12}-\left(\frac{\sigma_{\textrm{LJ}}}{r_{ij}}\right)^{6}\right] \ ,
\end{equation}
where $r_{ij}=||\mathbf{x}_i-\mathbf{x}_j||$ is the distance between two particles, $\sigma_{\textrm{LJ}}$ is the potential width (distance at which the potential is zero), and $\epsilon_{\textrm{LJ}}=30 \ k_\textrm{B}T$ is the depth of the potential well estimated from the experiments in Ref. \cite{dias_environmental_2023} (Figure \ref{fig:lj_fit}). For passive particles, we consider a truncated Lennard-Jones potential with a cut-off at $r_{\textrm{cut}}=2R = 2^{1/6}\sigma_{\rm LJ}$ to only consider a short-range steric repulsion (no attraction). For active particles, we also consider attraction with a cut-off set at $r_{\textrm{cut}}= 10R$ \cite{dias_environmental_2023}. 

The effective torque in Eq. \ref{eq:motion}b, which captures the feedback from the passive particles on the active particle's reorientation, 
is given by \cite{dias_environmental_2023, liebchen_which_2019}
\begin{equation}\label{eq:torque}
    \boldsymbol{\Omega}_i = - 4 \Omega_0 R^2[\hat{\mathbf{n}}_i, 0]\times\sum_j^{N_\textrm{p}}\nabla_{[\mathbf{x}_i,0]}\frac{e^{-\kappa r_{ij}}}{r_{ij}} \ ,
\end{equation}
\noindent
where $\Omega_0$ is the torque strength and $\kappa=(R/8)^{-1}$ is its screening length based on typical experimental values \cite{liebchen_which_2019}. We set a cut-off radius of $r_\textrm{cut}^\Omega=8R$, above which thermal noise dominates. 

\subsection*{Translational and rotational dynamics of living crystals}

As in earlier investigations with active colloids \cite{ginot_aggregation-fragmentation_2018}, we also quantify the translational and rotational dynamics of our living crystals using the distributions of their translational and angular speeds (Figs. \ref{fig:homo_vels}b-c and  \ref{fig:heter_vels}b-c). We consider a crystal distinct for as long as its size remains constant, thus avoiding aggregation and fragmentation events from influencing the analysis of the crystals' dynamics. To quantify the translational dynamics of a living crystal, we measure the average displacement-based velocity $v_\textrm{D}$ of its center of mass (COM) 
\begin{equation}\label{eq:vels}
    v_\textrm{D} = \frac{\left\langle \sum_i^{N} \big( \mathbf{x}_i(t)-\mathbf{x}_i(t-\delta) \big)\right\rangle}{N\delta} 
\end{equation}

\noindent
where $\langle...\rangle$ stands for time average, $\mathbf{x}_i$ is the position of particle $i$ (out of $N \geq 2$) in the crystal at time $t$, and $\delta=5000\Delta \tau = 0.1 \textrm{ s}$. To estimate the crystal's angular velocity $\omega$, we use the Kabsch-Umeyama algorithm \cite{kabsch_solution_1976,kabsch_discussion_1978,umeyama_least-squares_1991}, giving 

\begin{equation}
    \omega=  \frac{\left\langle\theta(t)\right\rangle}{\delta} \ ,
\end{equation}

\noindent
where $\theta(t)$ is the rotation angle at time $t$ extracted from the optimal rigid-body rotation matrix returned by the algorithm \cite{kabsch_solution_1976,kabsch_discussion_1978,umeyama_least-squares_1991}. For the rotational dynamics, we also calculate the torque exerted on the crystal's center of mass (COM) by the self-propulsion of its constituent particles, 

\begin{equation}\label{eq:clust_torque}
    \mathbf{M} = \gamma_{\rm t} v\sum_{i=1}^{N}{\mathbf{r}}_i\times\hat{\mathbf{n}}_i \ ,
\end{equation}

\noindent 
where $\mathbf{r}_i$ is the vector connecting the COM to particle $i$. The total torque $\mathbf{M}$ can be further divided into contributions $\mathbf{M}_\textrm{P}$ and $\mathbf{M}_\textrm{C}$ from perimeter and core particles respectively. The normalized signed scalar torque,
\begin{equation}\label{eq:rot_align}
    \Phi = \frac{1}{N}\sum_{i=1}^{N}\Phi_i  = \frac{1}{N}\sum_{i=1}^{N}\frac{\mathbf{M}_i}{ ||\mathbf{M}_i ||} \cdot \hat{\mathbf{u}}_z = \frac{1}{N}\sum_{i=1}^{N}\left(\hat{\mathbf{r}}_i\times\hat{\mathbf{n}}_i\right) \cdot \hat{\mathbf{u}}_z \ ,
\end{equation}

\noindent where $\hat{\mathbf{r}}_i$ is the unit vector of $\mathbf{r}_i$ and $\hat{\mathbf{u}}_z$ is the unit vector orthogonal to the plane of the crystal, provides a crystal-size-independent measure of the overall level of alignment of the active particles' self-propulsion directions with respect to the crystal's COM. $\Phi \in [-1, 1]$ with -1 and 1 indicating alignment compatible with clockwise and counterclockwise rotations, respectively, and 0 parallel and anti-parallel alignment to the radial direction connecting each particle to the COM. 

\subsection*{Average direction of nearby obstacles $\varphi$}

For each crystal, the average direction of nearby obstacles is given by

\begin{equation}\label{eq:var_phi_c}
    \varphi = \sum_{i=1}^{N}\varphi_i \ ,
\end{equation}
where $\varphi_i$ is the average weighted direction of nearby obstacles for each particle $i$ in the crystal (of $N$ particles) relative to its self-propulsion direction $\hat{\mathbf{n}}_i$, calculated as the weighted mean angle (Fig. \ref{fig:clust_rot}b) 

\begin{equation}
    \varphi_i = \frac{\sum_k^{n_\textrm{p}}w_{ik}\varphi_{ik}}{\sum_k^{n_\textrm{p}}w_{ik}} \ ,
\end{equation}
\noindent
where $\varphi_{ik}\in(-\pi,\pi]$ is the angle between $\hat{\mathbf{n}}_i$ and the vector connecting the center of particle $i$ to its passive neighbor $k$ (of $n_{\rm p}$ obstacles within particle $i$'s interaction range). The weights $w_{ik} = x^{-1}_{ik}$ are given by the inverse separation distance between particles $i$ and $k$. This choice ensures that obstacles closer to an active particle contribute more to $\varphi_i$, reflecting the inverse distance dependence of the torque $\mathbf{\Omega}_i$ (Eq. \ref{eq:torque}).

\subsection*{Delay from correlation}

We quantify the delay between any two signals $x$ and $y$ from the correlation $x*y$ between their time series. The correlation is obtained using SciPy \cite{scipy} (Version 1.16.3) function \texttt{signal.correlate} that calculates the correlation of the two signals by shifting $y$ backwards (or forwards) in time. The position of the maximum in the correlation at negative (positive) times means that $x$ precedes (follows) $y$. For example, we extract the delay $\tau_\textrm{delay}$ between the time evolution of the average obstacle direction $\varphi$ (Eq. \ref{eq:var_phi_c}) for a living crystal and that of the signed scalar torque $M$ on it (Eq. \ref{eq:clust_torque}) from the correlation $\varphi*M$ (Fig. \ref{fig:corr}). The position of the maximum in the correlation typically occurs at negative lag times, indicating that changes in $\varphi$ precede changes in $M$ (Fig. \ref{fig:corr}). Similarly, other delays in Fig. \ref{fig:clust_rev}e are extracted from the correlation between the two signals of interest. 

\subsection*{Radial moment $G$}

We calculate the radial moment $G$ of a living crystal as 

\begin{equation}
    G = \sum_{i=1}^{N} G_i = \gamma_\textrm{t} v \sum_{i=1}^{N} {\mathbf{r}}_i\cdot\hat{\mathbf{n}}_i\ .
\end{equation}

This quantity measures the overall tendency of the living crystal to expansion (positive $G$) or contraction (negative $G$) depending on whether particles tend to self-propel away or towards the crystal's COM on average. $G$ can be further divided into contributions $G_\textrm{P}$ and $G_\textrm{C}$ from perimeter and core particles, respectively. 

\subsection*{Phase shifts}

For a spinning living crystal, we extract the instantaneous phase shifts $\Delta\theta_\textrm{P}$ ($\Delta\theta_\textrm{C}$) between $|M|$ and $|M_\textrm{P}|$ ($|M_\textrm{C}|$) by first taking the Hilbert transform to obtain the respective analytic signals and then calculating the phase of the product of the analytic signal of $M$ and the complex conjugate of the analytic signal of $M_\textrm{P}$ ($M_\textrm{C}$) \cite{oppenheim_discrete_1999}. The average phase shift $\langle \Delta\theta_\textrm{P} \rangle$ ($\langle \Delta\theta_\textrm{C} \rangle$) is the average of these instantaneous values. \\

\medskip
\textbf{Acknowledgements} \par 
MPB and GV are grateful to the studentship funded by the A*STAR-UCL Research Attachment Programme through the EPSRC M3S CDT (EP/L015862/1). NAMA acknowledges financial support from the Portuguese Foundation for Science and Technology (FCT) under the contracts UID/PRR2/00618/2025 (https://doi.org/10.54499/UID/PRR2/00618/2025), UID/PRR/00618/2025 (https://doi.org/10.54499/UID/PRR/00618/2025), and UID/00618/2025 (https://doi.org/10.54499/UID/00618/2025). NAMA and GV also acknowledge support from the UCL MAPS Faculty Visiting Fellowship programme.

\medskip

\printbibliography

\newpage

\medskip
\textbf{Supplementary Figures} \par 

\renewcommand{\thefigure}{S\arabic{figure}}
\setcounter{figure}{0}

\begin{figure}[h]
  \centering
  \includegraphics[width=0.8\linewidth]{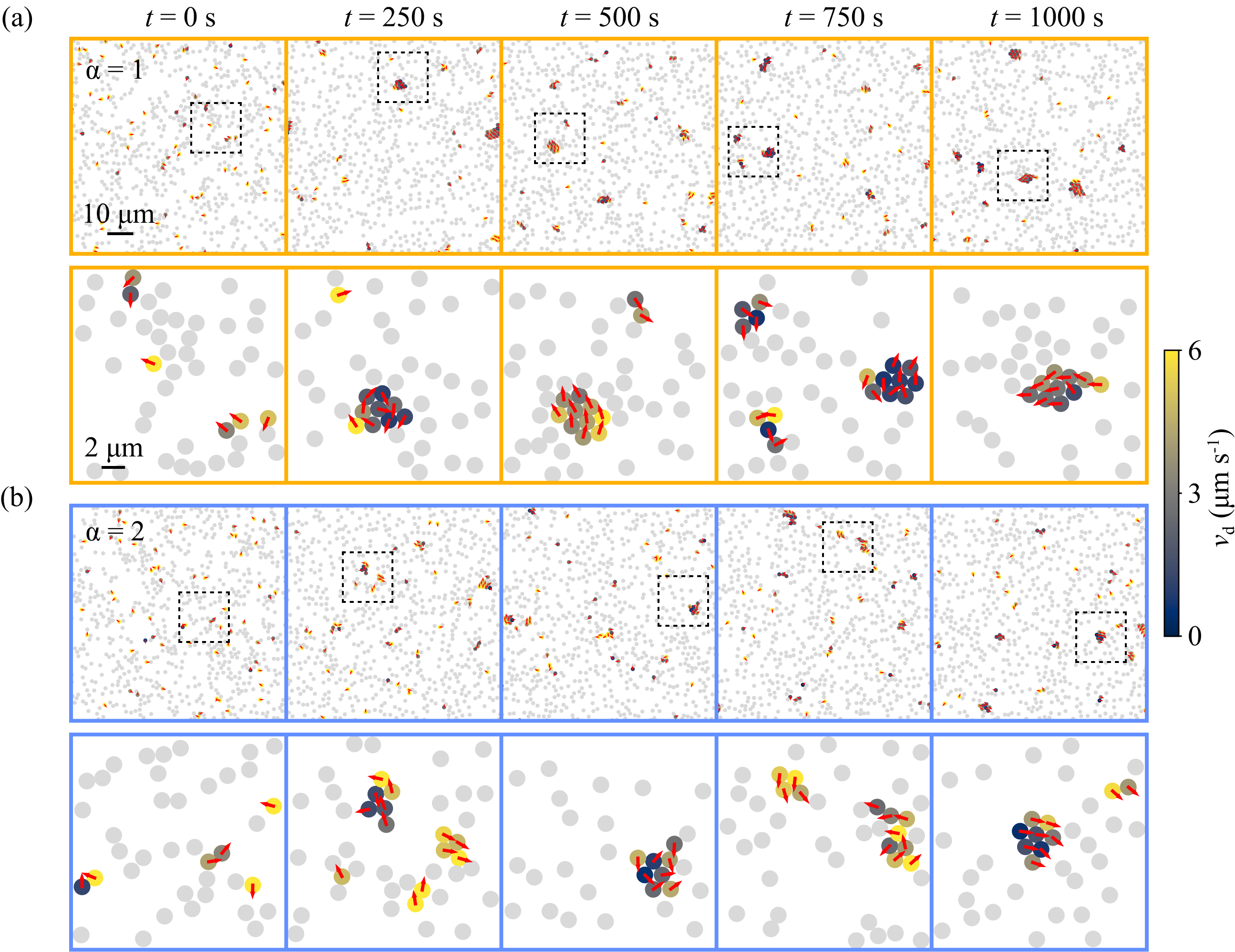}
  \caption{\textbf{Time evolution of the living crystals in Fig. \ref{fig:model_scheme}b-c at $\Omega_0 = 0 \ k_\textrm{B}T$}. Exemplary time evolutions of systems of run-and-tumble particles (colored-coded for their displacement speed $v_{\rm d}$ as in Fig. \ref{fig:model_scheme}b-c) in a representative crowded environment (gray, passive density $\rho_{\rm p}=15\%$) with $\Omega_0 = 0 \ k_\textrm{B}T$ for (a) $\alpha = 1$ (superdiffusion, Fig. \ref{fig:model_scheme}b) and (b) $\alpha = 2$ (normal diffusion, Fig. \ref{fig:model_scheme}c).  As the system evolves, living crystals form but remain relatively small as shown in the zooms in the bottom rows for selected areas in the respective top rows (dashed squares). 
  }
  \label{fig:clust_snaps_small}
\end{figure}

\begin{figure}[h]
  \centering
  \includegraphics[width=0.8\linewidth]{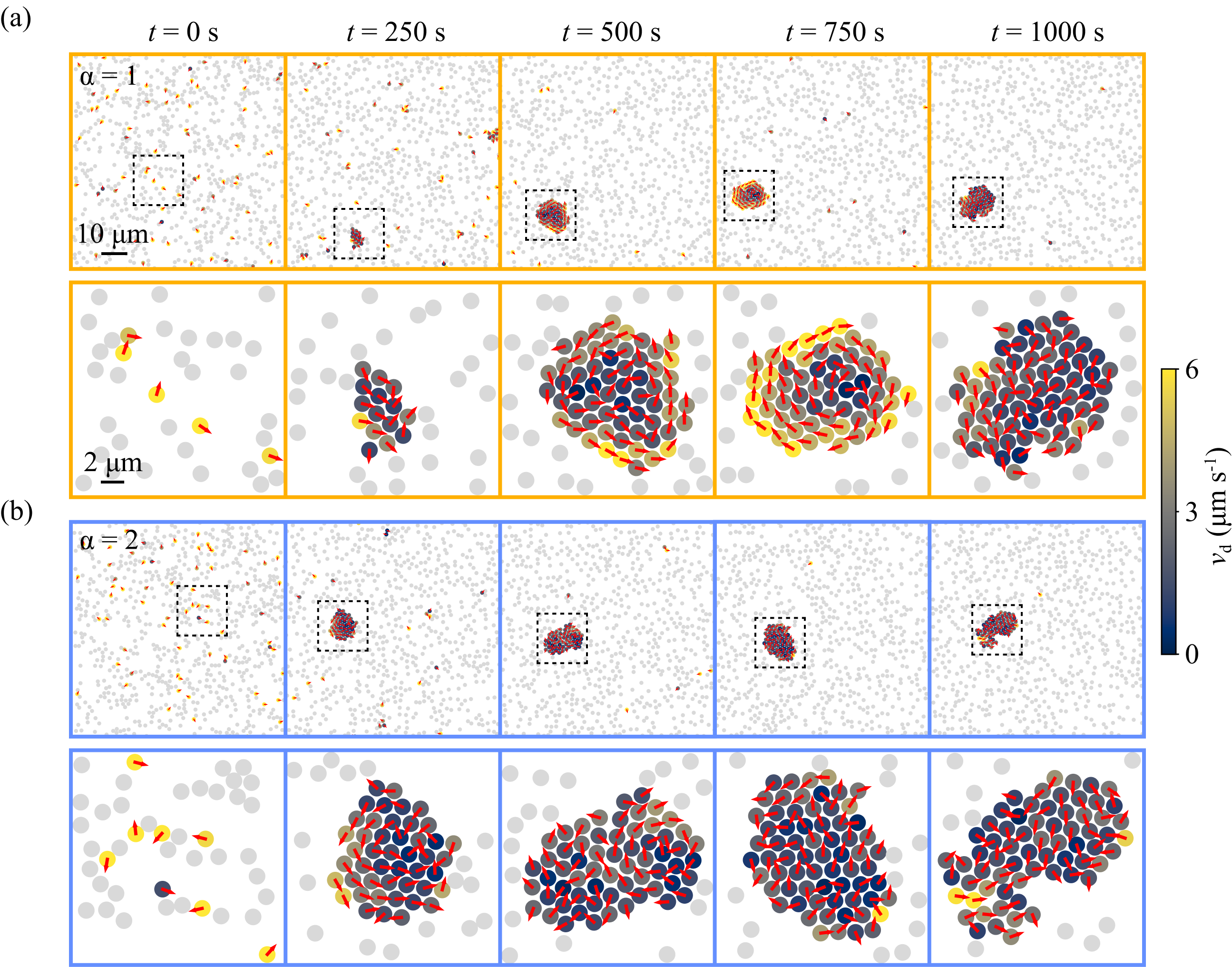}
  \caption{\textbf{Time evolution of the living crystals in Fig. \ref{fig:model_scheme}b-c at $\Omega_0 = 3 \ k_\textrm{B}T$}. Exemplary time evolutions of systems of run-and-tumble particles (colored-coded for their displacement speed $v_{\rm d}$ as in Fig. \ref{fig:model_scheme}b-c) in a representative crowded environment (gray, passive density $\rho_{\rm p}=15\%$) with $\Omega_0 = 3 \ k_\textrm{B}T$ for  (a) $\alpha = 1$ (superdiffusion, Fig. \ref{fig:model_scheme}b) and (b) $\alpha = 2$ (normal diffusion, Fig. \ref{fig:model_scheme}c). In contrast to  $\Omega_0 = 0 \ k_\textrm{B}T$ (Fig. \ref{fig:clust_snaps_small}), living crystals grow larger in size as shown in the zooms in the bottom rows for selected areas in the respective top rows (dashed squares).}
  \label{fig:clust_snaps_big}
\end{figure}

\begin{figure}[h]
  \centering
  \includegraphics[width=1\linewidth]{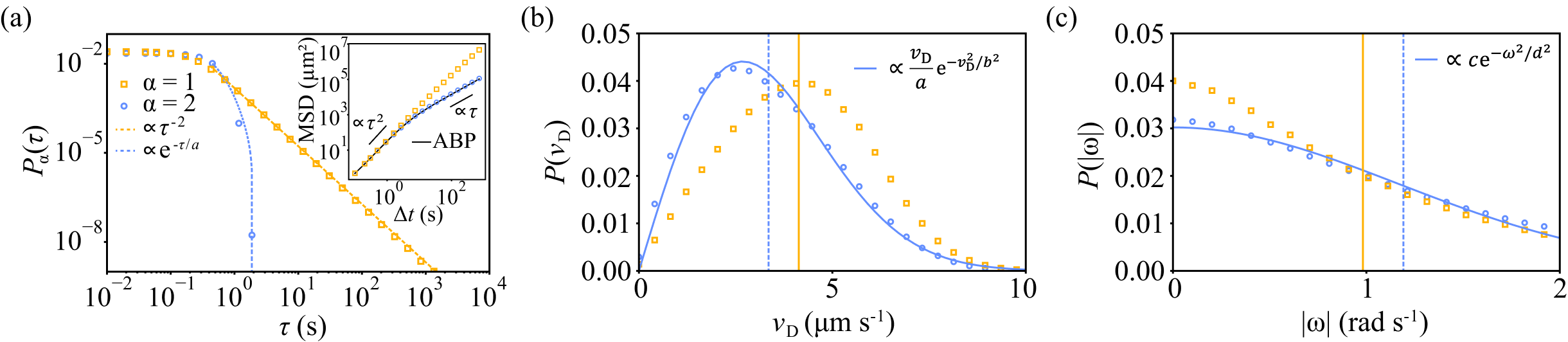}
  \caption{\textbf{Dynamics in homogeneous environments.} (a) The active particles' run durations are drawn from $\alpha$-stable L\'evy distributions $P_{\alpha}(\tau)$ for $\alpha = 2$ (blue circles) and  $\alpha = 1$ (yellow squares), yielding mean squared displacements (MSD) for individual agents that are asymptotically linear and ballistic (inset, diffusive and ballistic slopes shown for reference) in a homogeneous environment ($\rho_\textrm{p}=0\%$), respectively. The tails of the distributions scale as an exponential for $\alpha = 2$ and a power law $\propto \tau^{-2}$ for $\alpha = 1$, as confirmed by the respective fit curves (dashed lines). Both distributions were scaled by the same factor to guarantee that the MSD of the normal diffusive particles ($\alpha = 2$, blue circles) matches that of standard active Brownian particles of same size and self-propulsion speed (inset, black solid line). MSDs are averaged from $10$ simulations each with 68 non-interacting active particles. (b-c) Distributions of (b) the average center-of-mass velocities $v_\textrm{D}$ (translational dynamics) and (c) the absolute values $|\omega|$ of the angular speeds $\omega$ (rotational dynamics) for all simulated living crystals in a homogeneous environment ($\rho_{\rm} = 0\%$) for both normal diffusive ($\alpha = 2$) and superdiffusive ($\alpha = 1$) particles (Methods). The vertical lines indicate the average values of the distributions for $\alpha = 2$ (blue, $\langle v_{\rm D} \rangle = 3.40 \ \si{\micro\meter\per\second}$, $\langle |\omega| \rangle = 0.90 \ \si{\radian\per\second}$) and $\alpha = 1$ (yellow, $\langle v_{\rm D} \rangle = 3.84 \ \si{\micro\meter\per\second}$, $\langle |\omega| \rangle = 0.75 \ \si{\radian\per\second}$). The distributions for the normal diffusive particles are Rayleigh for $v_{\rm D}$ and Gaussian for $\omega$ as shown by fit lines (solid blue lines), consistent with experimental observations of phoretic colloids \cite{ginot_aggregation-fragmentation_2018}. All data obtained from time averages of size-constant crystals across 30 simulations lasting $1000 \textrm{ s}$.
}
  \label{fig:homo_vels}
\end{figure}
\begin{figure}[h]
  \centering
  \includegraphics[width=0.5\linewidth]{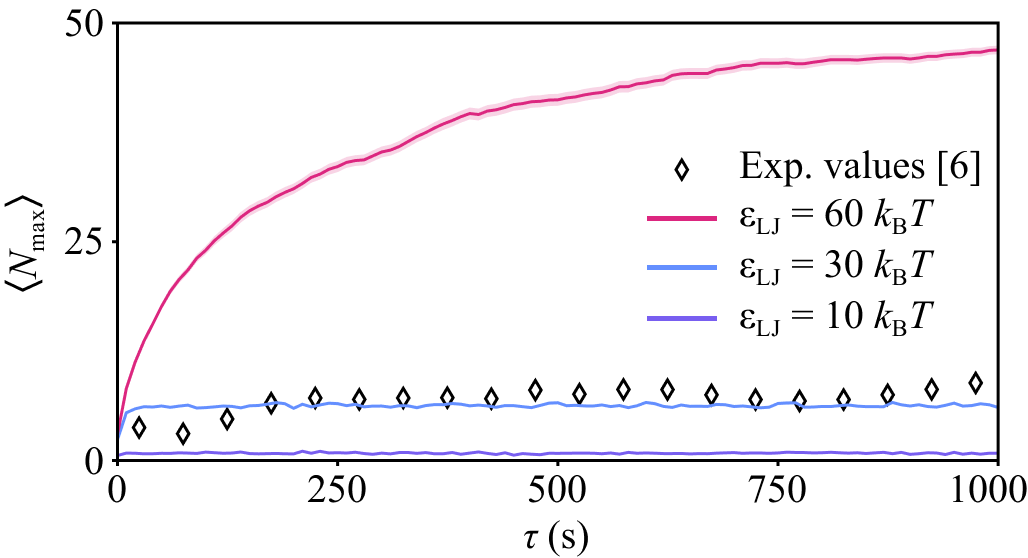}
  \caption{\textbf{Interaction potential between run-and-tumble particles.} Run-and-tumble particles interact with each other through a Lennard-Jones potential (Eq. \ref{eq:lj}, Methods) \cite{mognetti2013living}. We obtain a realistic estimate for the potential's well depth $\epsilon_{\rm LJ}$ between particles by tracking the average size of the largest crystal $\langle N_\textrm{max}\rangle$ as it evolves in time in a homogeneous environment ($\rho_\textrm{p} = 0 \%$) at an active density $\rho_\textrm{a}=1.1\%$ as in Ref. \cite{dias_environmental_2023} and by comparing it to experimental values (diamonds) for active colloids from the same Ref. \cite{dias_environmental_2023}. The systems' evolution is best represented by $\epsilon_\textrm{LJ}= 30 \ k_\textrm{B}T$ (blue solid line). All data are averaged across 200 independent simulations with shaded regions representing standard error (comparable to data points).
  }
  \label{fig:lj_fit}
\end{figure}

\begin{figure}[h]
  \centering
  \includegraphics[width=0.8\linewidth]{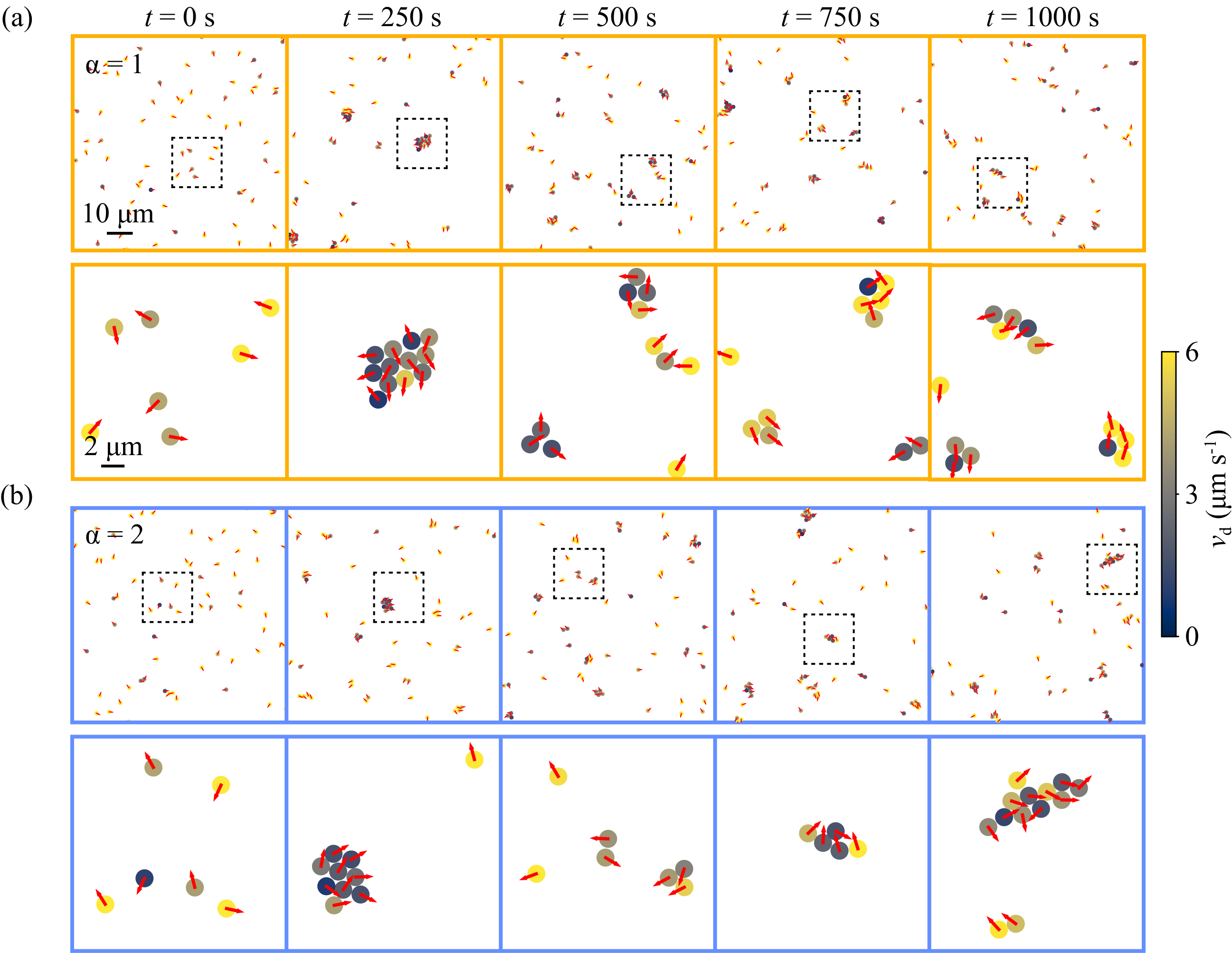}
  \caption{\textbf{Time evolution of  living crystals in homogeneous environments}.  Exemplary time evolutions of systems of run-and-tumble particles (colored-coded for their displacement speed $v_{\rm d}$) in the absence of obstacles ($\rho_{\rm p}=0\%$) for (a) $\alpha = 1$ (superdiffusion) and (b) $\alpha = 2$ (normal diffusion).  As the systems evolve, living crystals form but remain relatively small ($\langle N\rangle \approx5$, Fig. \ref{fig:clust_size}a) as shown in the zooms in the bottom rows for selected areas in the respective top rows (dashed squares). Living crystals of superdiffusive particles ($\alpha = 1$) are about 10\% larger than those of normal diffusive particles ($\alpha = 2$) due to a more efficient space exploration by individual particles before joining a crystal (MSDs in Fig. \ref{fig:homo_vels}a). Because of their longer average persistent time, superdiffusive particles also maintain alignment longer than their diffusive counterparts and form small solid-like living crystals that, on average, translate $\approx 23 \%$ faster (average center-of-mass speed $\langle v_{\rm D} \rangle = 4.13 \ \si{\micro\meter\per\second}$) but rotate $\approx 18\%$ slower (average absolute angular speed $\langle |\omega| \rangle = 0.98 \ \si{\radian\per\second} $) (Fig. \ref{fig:homo_vels}b-c). Qualitatively, these dynamics are  similar to those in heterogeneous environments in the absence of the environmental torque ($\Omega_0 = 0 \ k_\textrm{B}T$), where small living crystals of superdiffusive particles translate slightly faster ($\approx 16\%$) and rotate slightly slower ($\approx 15\%$) than those of normal diffusive particles  (Fig. \ref{fig:heter_vels}b-c).
  }
  \label{fig:clust_snaps_homogen}
\end{figure}

\begin{figure}[h]
  \centering
  \includegraphics[width=1\linewidth]{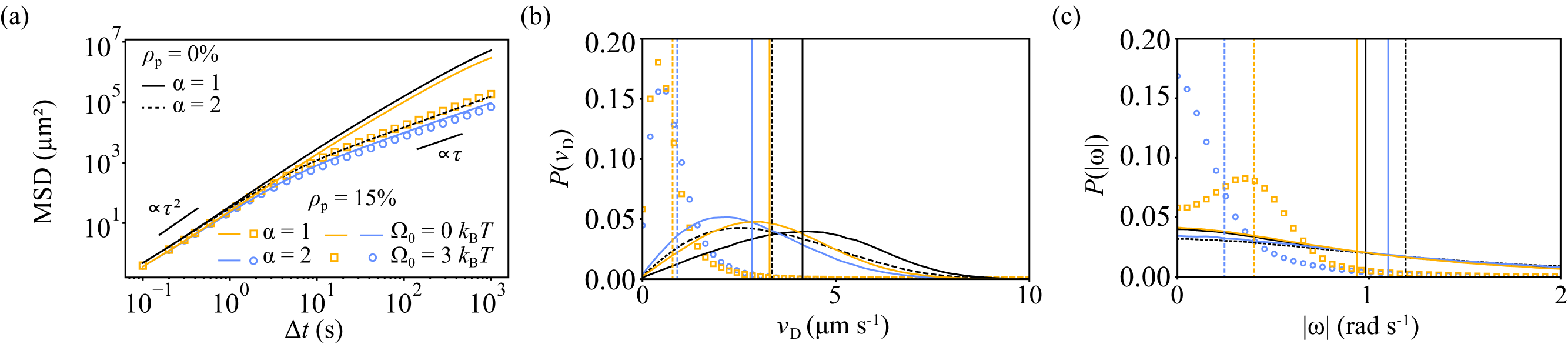}
  \caption{\textbf{Dynamics in heterogeneous environments.} (a) Mean squared displacements (MSD) of individual run-and-tumble particles, (b-c) distributions of (b) the average center-of-mass velocities $v_\textrm{D}$ (translational dynamics) and (c) the absolute values $|\omega|$ of the angular speeds $\omega$ (rotational dynamics) of living crystals in a representative heterogeneous environment ($\rho_\textrm{p} = 15\%$) for $\alpha = 1$ (yellow) and $\alpha = 2$ (blue) when $\Omega_0 = 0 \ k_\textrm{B}T$ (colored lines) and $\Omega_0 = 3 \ k_\textrm{B}T$ (colored symbols). In (a), MSDs are averaged from 10 simulations each with 68 non-interacting active particles, and diffusive and ballistic slopes are  shown for reference. In (b-c), the vertical lines indicate the average values of the respective distributions with (dashed lines) and without (solid lines) torque. The MSDs and distributions with respective averages in a homogeneous environment (Fig. \ref{fig:homo_vels}) are also shown for reference (black solid and dashed lines for $\alpha = 1$ and $\alpha = 2$, respectively). All crystal data obtained from 30 simulations lasting $1000$ s.
  }
  \label{fig:heter_vels}
\end{figure}

\begin{figure}[h]
  \centering
  \includegraphics[width=0.5\linewidth]{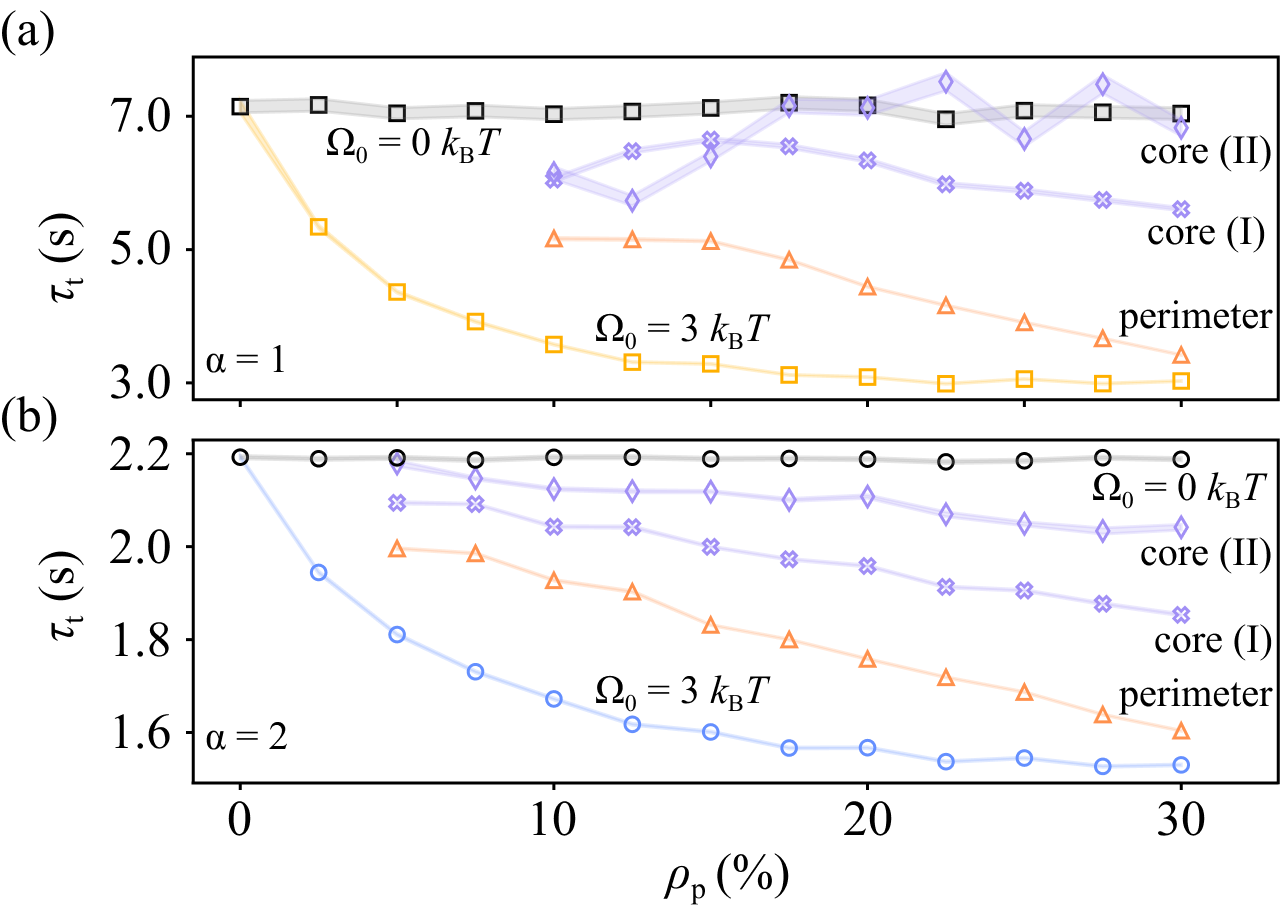}
  \caption{\textbf{Reorientation time $\tau_\textrm{t}$ of run-and-tumble particles in living crystals.} 
  Reorientation time $\tau_\textrm{t}$ of run-and-tumble particles based on their position within a living crystal (from perimeter, orange triangles, to core, purple symbols) as a function of passive density $\rho_\textrm{p}$ for different values of the L\'evy exponent $\alpha$ and strength $\Omega_0$ of the environmental torque: (a) $\alpha = 1$ (squares) and (b) $\alpha = 2$ (circles) at $\Omega_0 = 0 \ k_\textrm{B}T$ (gray symbols) and $\Omega_0 = 3 \ k_\textrm{B}T$ (yellow and blue symbols). Core particles are further divided based on their distance from the crystal's perimeter as in Fig. \ref{fig:clust_size}d: (I, purple crosses) $r_\textrm{P}=2R$ and (II, purple diamonds) $r_\textrm{P}\geq4R$. The environmental torque (Eq. \ref{eq:torque})  is strongest for perimeter particles and weakens for core particles toward the center of mass (COM), thus increasing $\tau_\textrm{t}$ from values characteristic of individual particles with torque (yellow squares for $\alpha = 1$ and blue circles for $\alpha = 2$, as in Fig. \ref{fig:clust_size}b) at the perimeter to those characteristic of individual particles without torque (gray squares for $\alpha = 1$ and gray circles for $\alpha = 2$, as in Fig. \ref{fig:clust_size}b) near the COM at each $\rho_{\rm p}$ ($\rho_{\rm p} = 15\%$ as in Fig. \ref{fig:clust_size}d). Overall, $\tau_\textrm{t}$ tends to reduce with increasing $\rho_\textrm{p}$, pointing at an increased crystal stabilization. The non-monotonic trend observed for core particles for $\alpha = 1$ is a reflection of the non-monotonic trend in crystal size (Fig. \ref{fig:clust_size}a), as particles within the crystal are more shielded from interactions with obstacles for larger living crystals. This effect is more prominent for $\alpha = 1$ than for $\alpha = 2$ as peak crystal formation is shifted towards a larger $\rho_{\rm p}$ (Fig. \ref{fig:clust_size}a). Data obtained from crystals with at least 20 particles. Shaded areas represent standard error and are typically smaller than the symbol size.
 }
  \label{fig:tau_t_clust}
\end{figure}

\begin{figure}[h]
  \centering
  \includegraphics[width=0.8\linewidth]{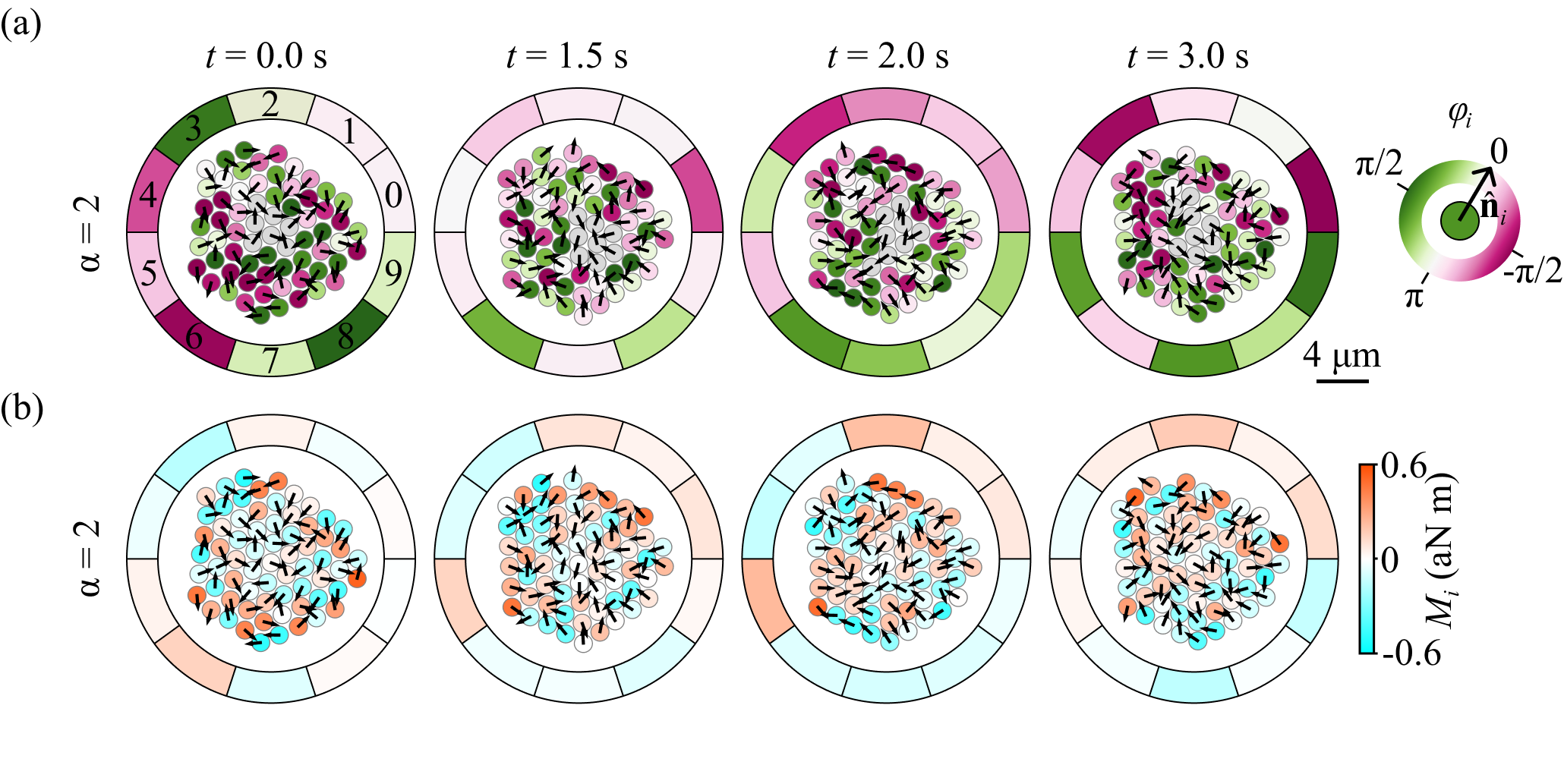}
  \caption{\textbf{Exemplary rotational dynamics of diffusive living crystals ($\alpha = 2$).} Exemplary sequence of a standard diffusive living crystal ($\alpha = 2$) showing (a) the average weighted direction of passive obstacles $\varphi_{i}$ and (b) the contribution $M_i$ to the signed scalar torque $M$ on the crystal (Methods) for every run-and-tumble particle $i$ within the living crystal. Positive and negative values are more evenly distributed within the crystal and over time compared to the sequence for $\alpha = 1$ in Fig. \ref{fig:clust_rot}c-d. Color scales as in Fig. \ref{fig:clust_rot}. Black arrows represent the unit vectors $\hat{\mathbf{n}}_i$ of each particle's self-propulsion direction. Outer annular arcs (from 0 to 9) show the sector average of the respective quantity. Core gray particles in (a) do not interact with obstacles.
}
  \label{fig:a2_snaps}
\end{figure}

\begin{figure}[h]
  \centering
  \includegraphics[width=0.8\linewidth]{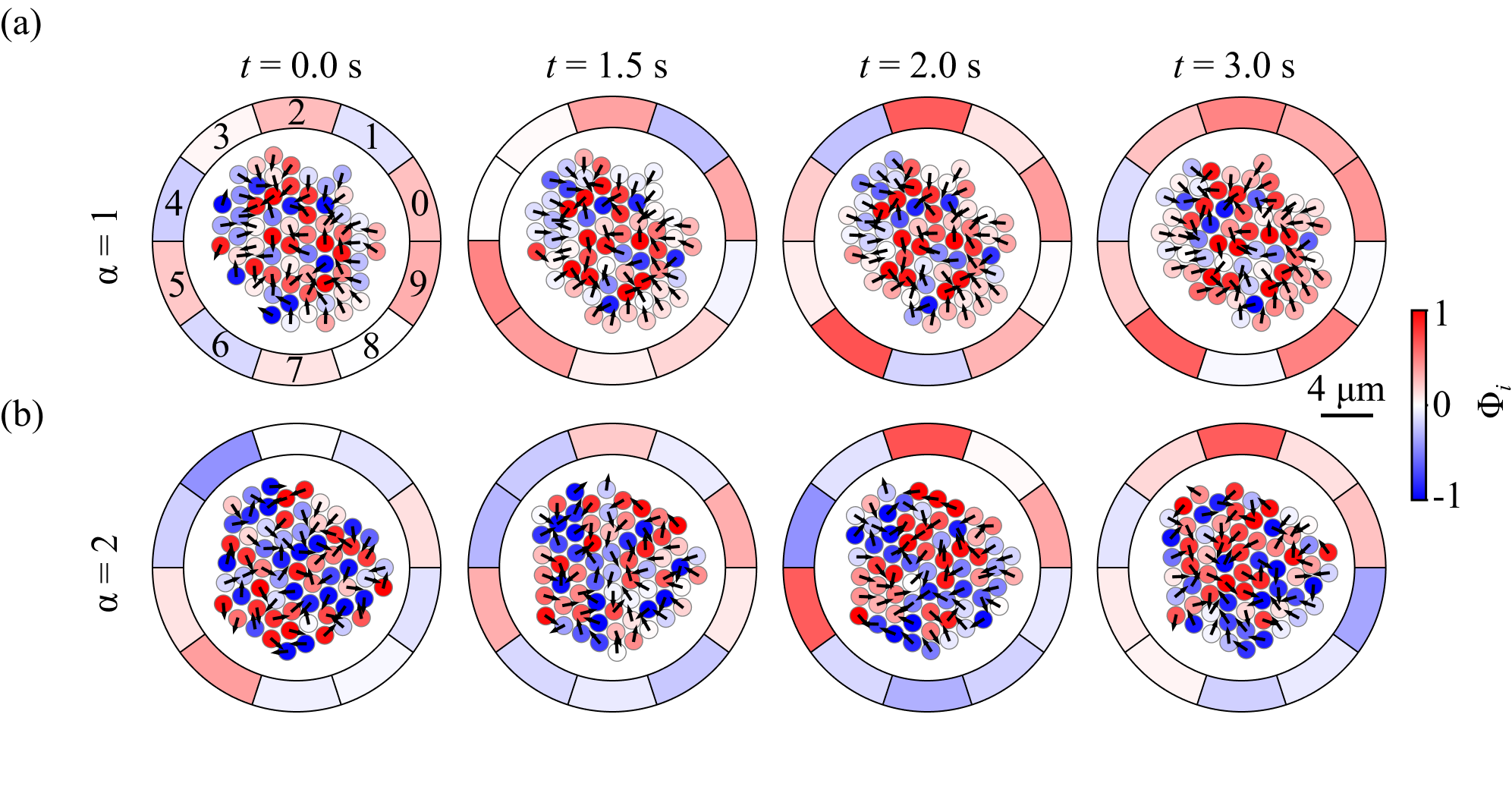}
  \caption{\textbf{Normalized signed scalar torque $\Phi$ in living crystals.}  The normalized scalar torque $\Phi_i$ (Methods) for each run-and-tumble particle $i$ in the exemplary sequences of (a) Fig. \ref{fig:clust_rot}c-d for $\alpha = 1$ and (b) Fig. \ref{fig:a2_snaps} for $\alpha = 2$. The values show particle alignment compatible with a counterclockwise rotation emerging for $\alpha = 1$ (predominant red) but no clear alignment emerging for $\alpha = 2$ due to more frequent tumbling events. Black arrows represent the unit vectors $\hat{\mathbf{n}}_i$ of each particle's self-propulsion direction. Outer annular arcs (from 0 to 9) show the sector average of $\Phi_i$. 
}
  \label{fig:rot_order}
\end{figure}

\begin{figure}[h]
  \centering
  \includegraphics[width=1\linewidth]{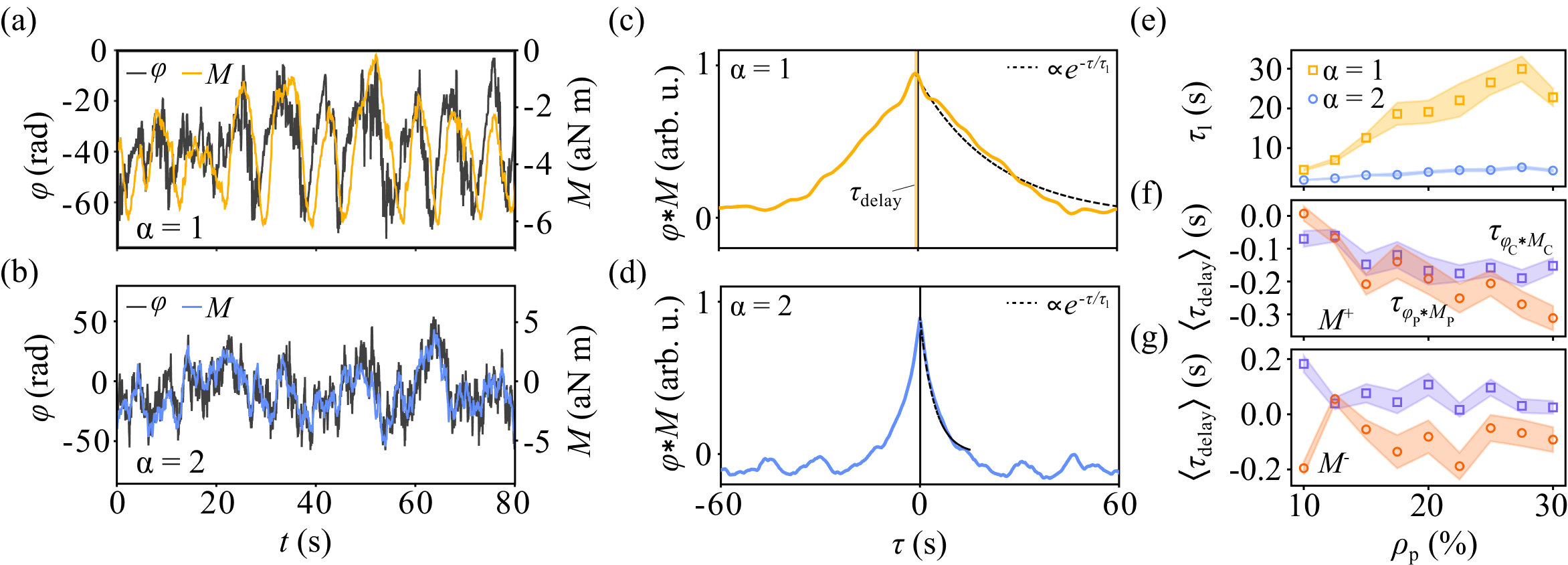}
  \caption{\textbf{Correlation between $\varphi$ and $M$}. (a-b) Exemplary signals of the average direction $\varphi$ of nearby obstacles (black line) and signed scalar torque $M$ (colored lines) for two living crystals with (a) $\alpha = 1$ and (b) $\alpha = 2$ at $\rho_{\rm p} = 15 \%$. For $\alpha = 1$, a delay between the two is visible, with $M$ following $\varphi$. (c-d) Normalized correlations $\varphi*M$ between the two signals for (c) $\alpha = 1$ and (d) $\alpha = 2$.  The position of the peak in the correlation (colored vertical line) measures the time delay $\tau_\textrm{delay}$ between the two signals. The correlation is peaked at a slightly negative value for $\alpha = 1$, while it is peaked closer to zero (black vertical lines) for $\alpha = 2$. (e) The decay $\tau_\textrm{l}$ of the correlations is obtained by fitting to $e^{-\tau/\tau_\textrm{l}}$ as exemplified in c-d (dashed line) and shown at different $\rho_\textrm{p}$ values for $\alpha = 1$ (yellow) and $\alpha = 2$ (blue). The correlation decays much faster for $\alpha = 2$ than for $\alpha =1$. (f-g) Values of $\langle \tau_\textrm{delay} \rangle$ in Fig. \ref{fig:clust_rot}f for $\alpha = 1$ divided by the contribution from perimeter (orange circles, $\varphi_{\rm P}*M_{\rm P}$) and core (purple squares, $\varphi_{\rm C}*M_{\rm C}$) particles during periods of (f) speeding up ($M^+$) and (g) slowing down ($M^-$) in crystal rotation. Data in e-g obtained as averages across 60 independent simulations lasting 1000 s; shaded regions represent standard error.
  }
  \label{fig:corr}
\end{figure}

\begin{figure}[h]
  \centering
  \includegraphics[width=0.7\linewidth]{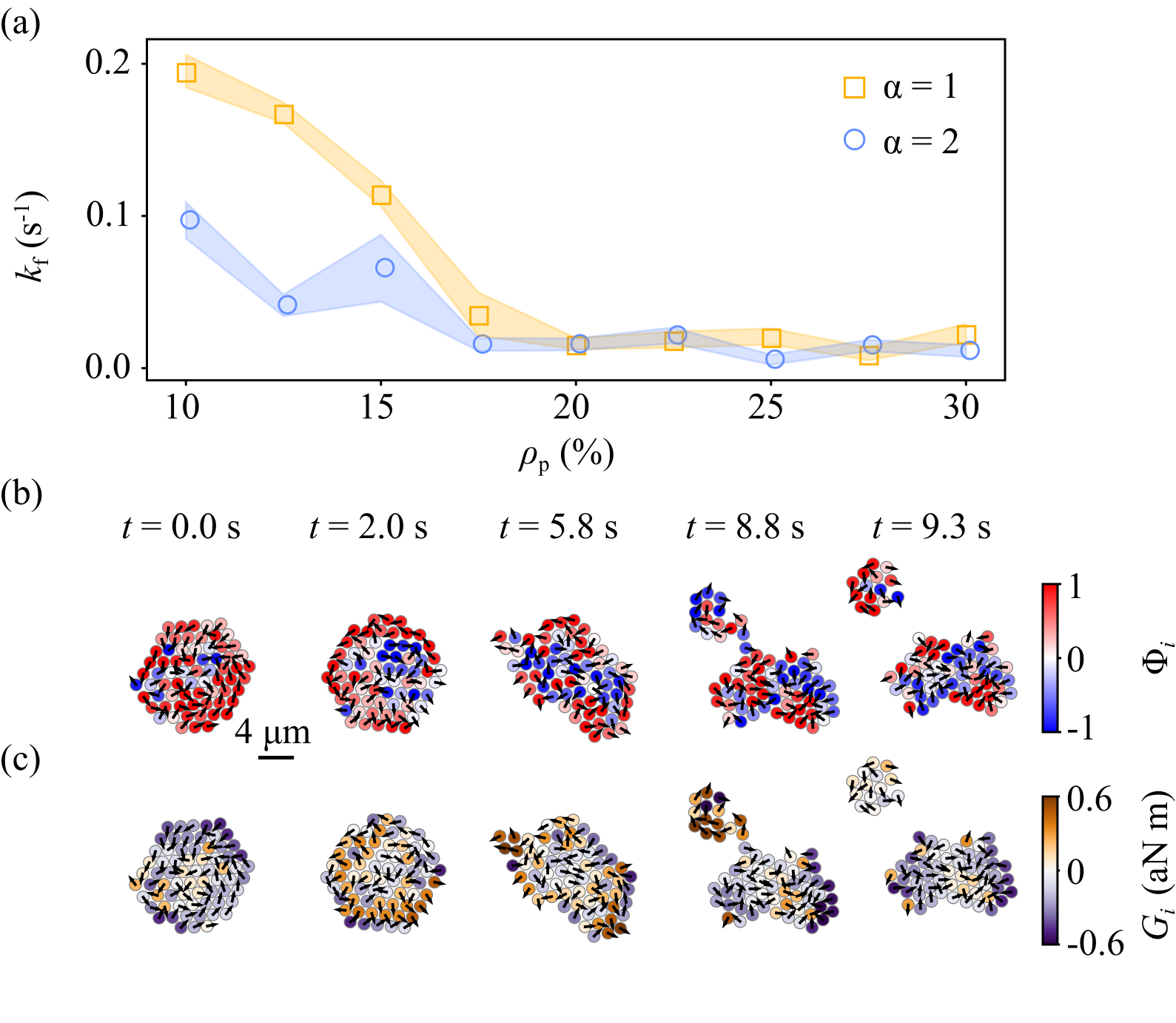}
  \caption{\textbf{Spinning-induced fragmentation of living crystals at low obstacle densities.} (a) Rate $k_\textrm{f}$ of living crystal fragmentation for superdiffusive ($\alpha = 1$, yellow) and normal diffusive particles ($\alpha = 2$, blue). The rate drops to very low values $k_\textrm{f}\approx 0.02 \,{\rm s^{-1}}$ at high passive densities $\rho_{\rm p} \ge 17.5\%$ due to the increased stabilizing effect of the environmental torque (Fig. \ref{fig:clust_size}). At lower $\rho_{\rm p}$, fragmentation occurs more often for $\alpha = 1$ due to their higher persistence time. $k_\textrm{f}$ is obtained by counting the number of fragmentation events where a crystal loses more than 2 particles in a single timestep. The measurement is obtained as an average across 60 simulations lasting 1000 s for each value of $\alpha$; shaded regions represent standard error. (b) Normalized signed scalar torque $\Phi_i$ and (c) radial moment $G_i$ for each particle $i$ (Methods) in an exemplary sequence of spinning-induced fragmentation of a living crystal for $\alpha = $1 at  $\rho_{\rm p} =15 \%$. Black arrows indicate $\hat{\mathbf{n}}_i$. The crystal rotates quickly ($t=0 \textrm{ s}$) causing some of the perimeter and core particle's $\hat{\mathbf{n}}_i$ to drift away from the COM ($t=2 \textrm{ s}$). If the environmental torque is too weak to stabilize the crystal and re-orient $\hat{\mathbf{n}}_i$ towards the COM fast enough within a spinning cycle, the crystal starts to elongate ($t=5.8 \textrm{ s}$) until it eventually fragments ($t\geq 5.8 \textrm{ s}$). Spinning induced fragmentation is suppressed with increasing environmental torque and becomes less frequent as $\rho_\textrm{p}$ increases.}
  \label{fig:clust_break}
\end{figure}

\begin{figure}[h]
  \centering
  \includegraphics[width=0.4\linewidth]{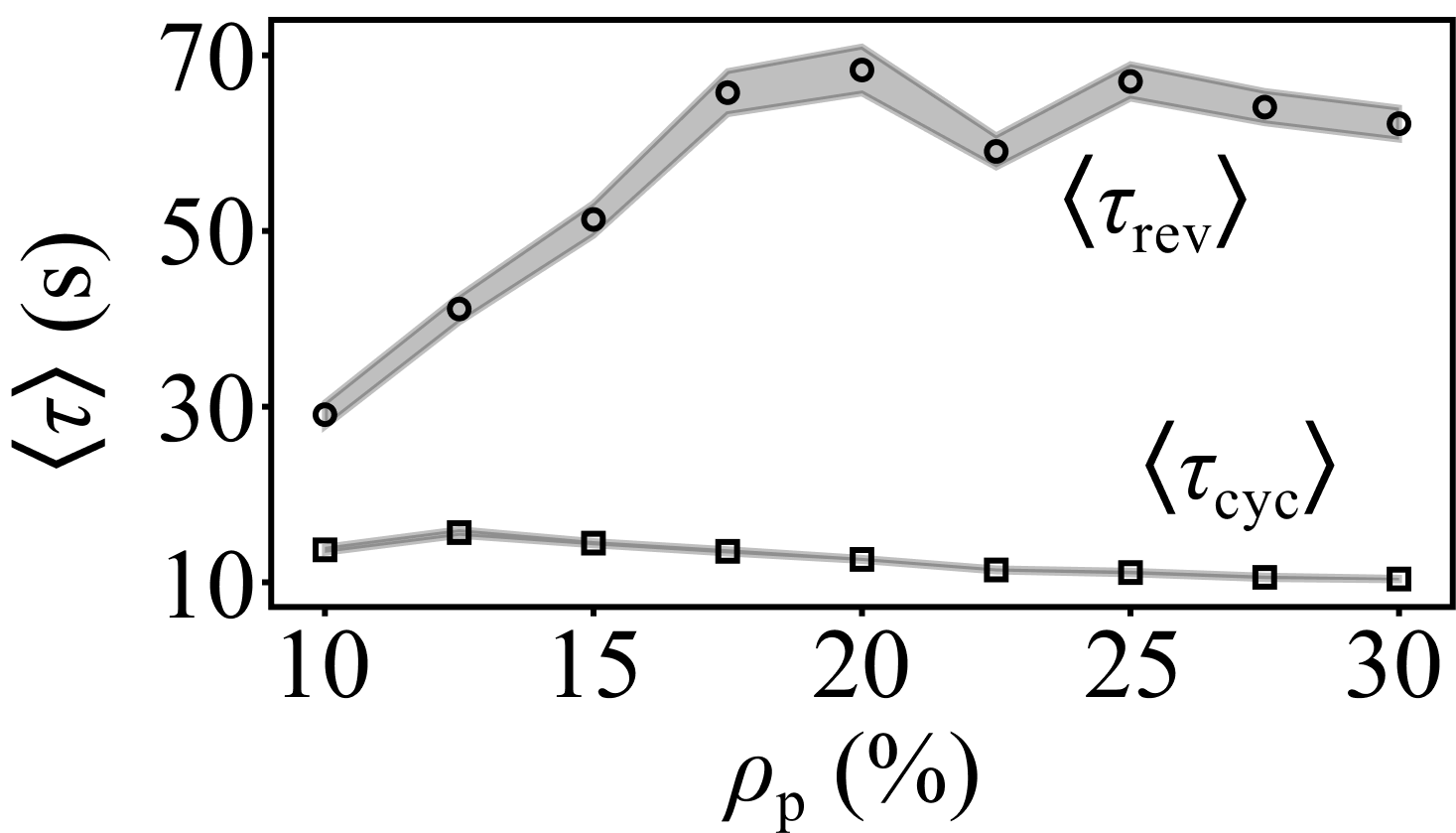}
  \caption{\textbf{Timescales of spinning living crystals' rotational dynamics.} Average times $\langle \tau_\textrm{cyc} \rangle$ (squares) of a spinning cycle (peak-to-peak time) and $\langle \tau_\textrm{rev} \rangle$ (circles) between reversal events in crystal rotation. The period of a spinning cycle is determined by the balance of two antagonistic effects with obstacle density $\rho_\textrm{p}$: the crystal's size and the strength of the environmental torque. As angular speed tends to decrease with size, $\langle \tau_\textrm{cyc} \rangle$ values follow a non-monotonic trend resembling that in Fig. \ref{fig:clust_size}; the peak is however shifted to smaller $\rho_\textrm{p}$ as stronger torques at higher obstacle densities induce faster particles' reorientation towards the crystal's COM, thus reducing $\langle \tau_\textrm{cyc} \rangle$. $\langle \tau_\textrm{rev} \rangle$ instead tends to increase with $\rho_\textrm{p}$ due to the weaker influence of environmental fluctuations at higher obstacle densities (Fig. \ref{fig:clust_rot}f), making reversal events less frequent. Data averaged across 60 independent simulations; shaded area represent standard error.
  }
  \label{fig:time_rev}
\end{figure}

\begin{figure}[h]
  \centering
  \includegraphics[width=0.7\linewidth]{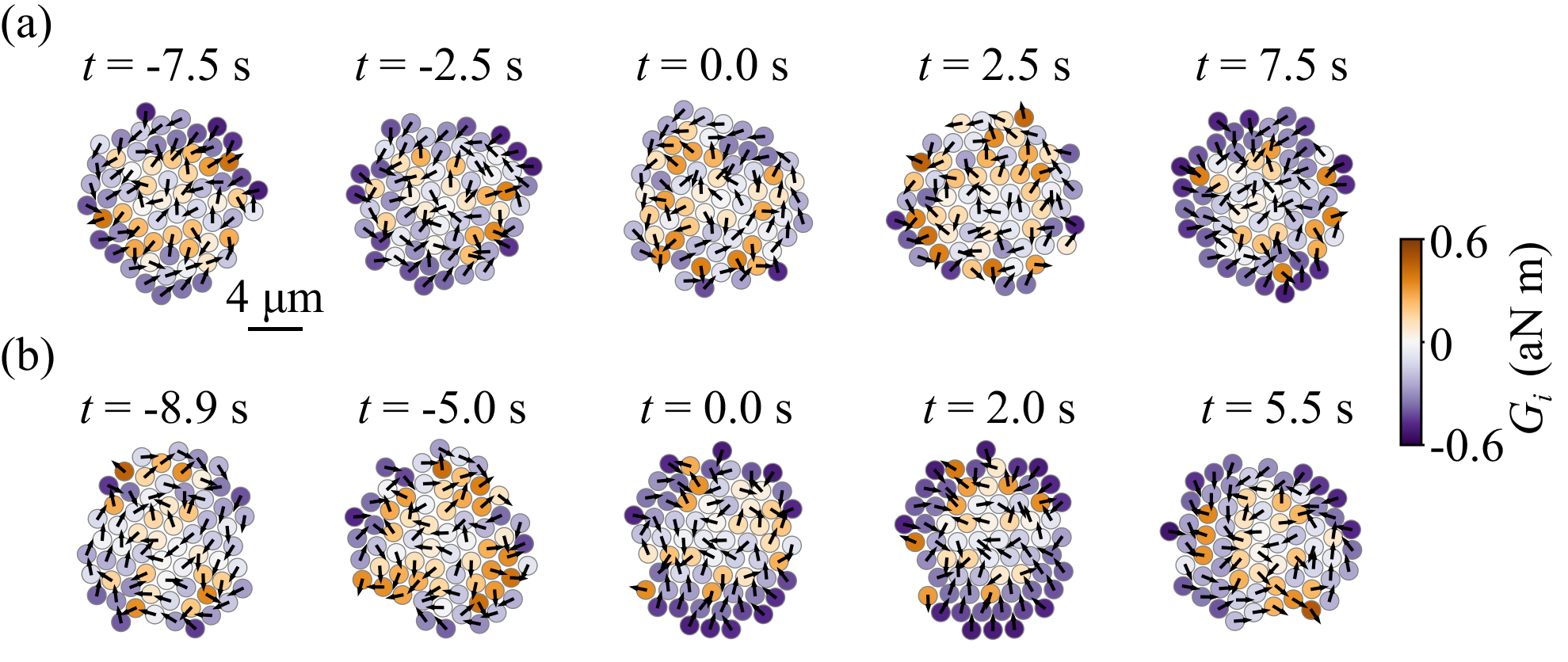}
  \caption{\textbf{Radial moment $G$ for the sequences in Fig. \ref{fig:clust_rev}.} The radial moment $G_i$ for each particle $i$ in the exemplary sequences of (a) Fig. \ref{fig:clust_rev}a for a crystal maintaining its direction of rotation and (b) Fig. \ref{fig:clust_rev}b for a crystal reversing its direction. Both crystals slow down due to an alignment of the core particles away from the respective crystal's COM (positive $G_i$): (a) from $t = 0 \, {\rm s}$ to $t = 2.5 \, {\rm s}$ and (b) from $t = -5 \, {\rm s}$ to $t = 0 \, {\rm s}$. Black arrows represent the unit vectors $\hat{\mathbf{n}}_i$ of each particle's self-propulsion direction.}
  \label{fig:viril_dyn}
\end{figure}

\end{document}